\documentclass[preprint]{elsarticle}

\usepackage[utf8x]{inputenc}
\usepackage{times}

\usepackage{amsmath,amsthm}
\theoremstyle{definition}
\newtheorem{definition}{Definition}
\newtheorem{example}{Example}

\usepackage{hyperref}

\newif\iflong
\longfalse

\usepackage{changepage} 
\usepackage[inline]{enumitem}

\usepackage{numprint}

\usepackage{booktabs}
\usepackage{relsize}

\makeatletter\let\expandableinput\@@input\makeatother

\usepackage{fontawesome}
\newcommand{\checkOK}[1][black]{\text{\color{#1}\faCheck}}

\newcommand{\checkQM}[1][black]{\text{\color{#1}\faQuestion}}

\newcommand{\fakepar}[1]{\textbf{#1}}

\graphicspath{{../pdf/}{../jpeg/}}
\DeclareGraphicsExtensions{.pdf,.jpeg,.png}
\usepackage{amsmath}

\usepackage{array}
\usepackage{listings}
\usepackage{multicol,multirow}
\usepackage[font=small]{caption}
\usepackage{subcaption}

\usepackage{tikz}
\usetikzlibrary{shapes,arrows,positioning,calc,fit,intersections,through,automata,matrix}

\definecolor{vgreen}{RGB}{27,158,119}
\definecolor{vorange}{RGB}{217,95,2}
\definecolor{vviolet}{RGB}{117,112,179}
\definecolor{vpink}{RGB}{231,138,195}

\usepackage{url}
\usepackage{xspace}

\usepackage[T1]{fontenc}
\usepackage[scaled=0.81]{beramono}  
\usepackage[plain, linesnumbered, noend]{algorithm2e}

\SetAlCapSkip{1.2em}
\SetInd{2mm}{1mm}
\SetKwIF{cIf}{cElseIf}{cElse}{if}{}{elseif}{else}{endif}

\newcommand{\toolname}{{\smaller[0.5]\textsc{Plumb\-Droid}}\xspace}
\newcommand{\J}[1]{{\mbox{\lstinline[basicstyle=\ttfamily]|#1|}}}  
\newcommand{\Jm}[1]{\text{\J{#1}}}  
\usepackage[colorinlistoftodos,textwidth=10mm,textsize=tiny,obeyFinal]{todonotes}

\newcommand{\rec}{\ensuremath{\leftarrow}}     
\newcommand{\var}[1]{\ensuremath{\mathit{#1}}} 

\newcommand{\lang}{\ensuremath{\mathcal{L}}}

\newcommand{\droidl}{{\smaller[0.5]\textsc{DroidLeaks}}\xspace}
\newcommand{\leaks}{\droidl}

\usepackage{boxedminipage}
\newenvironment{result}%
{\smallskip
	\noindent
	\let\emph=\textbf
	\begin{boxedminipage}{\columnwidth}\begin{center}\em}%
		{\end{center}\end{boxedminipage}%
	\smallskip
}

\SetArgSty{textnormal}

\everypar{\looseness=-1}

\newcommand{\nicepar}[1]{\textbf{#1}}

\usepackage{flushend}

\begin{document}

\title{Automated Repair of Resource Leaks\\ in Android Applications}

\author[1]{Bhargav Nagaraja Bhatt}
\ead{bhattb@usi.ch}

\author[1]{Carlo A. Furia}
\ead[url]{bugcounting.net}

\address[1]{Software Institute, USI Universit\`a della Svizzera italiana, Lugano, Switzerland}

\begin{abstract}
	Resource leaks---a program does not release resources it previously acquired---are
	a common kind 
	of bug in Android applications.
	Even with the help of existing techniques to automatically detect leaks,
	writing a leak-free program remains tricky.
	One of the reasons is Android's event-driven programming model, which complicates the understanding of an application's overall control flow.
	
	In this paper, we present \toolname: a technique to automatically \emph{detect and fix} resource leaks in Android applications.
	\toolname builds a succinct abstraction of an app's control flow, and uses it to find execution traces that may leak a resource.
	The information built during detection also enables automatically building a fix---consisting of release operations
	performed at appropriate locations---that removes the leak and does not otherwise affect the application's usage of the resource.
	
	An empirical evaluation on resource leaks from the \leaks curated collection
	demonstrates that \toolname's approach is scalable, precise,
	and produces correct fixes for a variety of
	resource leak bugs: 
	\toolname automatically found and repaired 50 leaks 
	that affect 9 widely used resources of the Android system,
	including all those collected by \leaks for those resources;
	on average, it took just 2 minutes to detect and repair a leak.
	\toolname also compares favorably to Relda2/RelFix---the only other fully automated approach to repair Android resource leaks---
	since it can often detect more leaks with higher precision and producing smaller fixes.
	These results indicate that \toolname
	can provide valuable support to enhance the quality of Android applications in practice.
\end{abstract}

\maketitle

\section{Introduction}

The programming model of the Android operating system
makes its mobile applications (``apps'')
prone to bugs that are due to incorrect usage of shared resources.
An app's implementation typically runs from several entry points, which are activated by callbacks of the Android system
in response to events triggered by the mobile device's user (for example, switching apps) or other changes in the environment
(for example, losing network connectivity).
Correctly managing shared resources is tricky in such an event-driven environment,
since an app's overall execution flow is not apparent from the control-flow structure of its source code.
This explains why \emph{resource leaks}---bugs that occur when a shared resource is not correctly released or released too late---%
are one of the most common kind of 
performance bugs in Android apps~\cite{LeakEmpericalStudy,icse14performancebugs,emseperformancebugs},
where they often result in buggy behavior that ultimately degrades an app's responsiveness and usability.

Motivated by their prevalence and negative impact, 
research in the last few years (which we summarize in \autoref{sec:relatedwork}) has
developed numerous techniques to \emph{detect} resource leaks
using dynamic analysis~\cite{LeakEmpericalStudy, FSE14}, static analysis~\cite{RountevEnergyBugs, Relda2TSE16, Relfix}, or a combination of both~\cite{EnergyPatchTSE18}.
Automated detection is very useful to help developers in debugging,
but the very same characteristics of Android programming that make apps prone to having resource leaks
also complicate the job of coming up with leak \emph{repairs} that are correct in all conditions.

To address these difficulties, we present a technique to detect \emph{and fix} resource leaks in Android apps completely automatically.
Our technique, called \toolname and described in \autoref{sec:how-it-works}, is based on static analysis 
and can build 
fixes that are correct (they eradicate the detected leaks for a certain resource)
and ``safe'' (they do not introduce conflicts with the rest of the app's usage of the resource,
and follow Android's recommendations for resource management~\cite{AndoidActivityLifecycleGuidelines}).

\toolname's analysis is scalable because it is based on a succinct abstraction of an app's
control-flow graph called \emph{resource-flow graph}.
Paths on an app's resource-flow graph correspond to all its possible usage of resources.
Avoiding leaks entails matching each acquisition of a resource with a corresponding release operation.
\toolname supports the most general case of reentrant resources (which can be acquired multiple times,
typically implemented with reference counting in Android):
absence of leaks is a context-free property~\cite{Sipser};
and leak detection amounts to checking 
whether every path on the resource-flow graph
belongs to the context-free language of leak-free sequences. 
\toolname's leak model is more general than most other leak detection techniques'---which
are typically limited to non-reentrant resources (see \autoref{tab:related-work}).

The information provided by our leak detection algorithm
also supports the automatic \emph{generation of fixes} that remove leaks.
\toolname builds fixes that are correct by construction;
a final validation step reruns the leak detection algorithm augmented with the property that
the new release operations introduced by the fix do not interfere with the existing resource usages.
Fixes that pass validation are thus correct and ``safe'' in this sense.

We implemented our technique \toolname in a tool, also called \toolname,
that works on Android bytecode.
\toolname can be configured to work on any Android resource API;
we equipped it with the information about acquire and release operations of 9 widely used Android resources (including \J{Camera} and \J{WifiManager}),
so that it can automatically repair leaks of those resources. 

The current implementation of \toolname
is not equipped with any aliasing analysis technique;
therefore, it may incur a large number of false positives
when applied to apps that use resources under lots of different aliases.
We found that different Android resources have distinct usage patterns:
some, such as database cursors, are frequently used under many aliases;
others, such as the camera or the Wi-Fi adapter, are not.
\toolname's current implementation is geared towards analyzing
the latter kind of resources, which we call \emph{non-aliasing} resources
and are the main target of our experiments with \toolname. 

We evaluated \toolname's performance empirically on leaks of 9 non-aliasing resources
in 17 Android apps 
from the curated collection DroidLeaks~\cite{droidleaks}.
These experiments, described in \autoref{sec:experiments},
confirm that \toolname is a scalable automated leak repair technique
(around 2 minutes on average to find and repair a leak)
that consistently produces correct and safe fixes for a variety of Android resources
(including all 26 leaks in \leaks affecting the 9 analyzed resources).

We also experimentally compared \toolname with Relda2/RelFix~\cite{Relda2TSE16,Relfix}---the only other fully automated approach
to repair Android resource leaks that has been developed so far---by running the former on the same apps used in the latter's
evaluation.
The comparison, also described in \autoref{sec:experiments},
indicates that, on non-aliasing resources,
\toolname detects more true leaks (79 vs.\ 53)
with a higher average precision (89\% vs.\ 55\%)
than Relda2/RelFix,
and produces fixes that are one order of magnitude smaller.

\toolname's novelty and effectiveness lie in how it combines
and fine-tunes existing analysis techniques to the usual 
characteristics of Android resource usage.
\begin{enumerate*}
	\item \toolname's fine-grained, yet succinct, abstract model of resource usage
	supports a sound static analysis with high precision;
	\item \toolname's fix generation process follows Android's guidelines about
	the program locations where each resource should be released;
	\item \toolname's validation step further guarantees that the fixes that it automatically generates
	are suitable and correct.
\end{enumerate*}

In summary, this paper makes the following contributions:
\begin{itemize}
	\item It introduces \toolname: a fully automated technique based on static analysis
	for the detection and repair of Android resource leaks.
	\item It evaluates the performance of \toolname on apps in \leaks,
	showing that it achieves high precision and recall, and
	scalable performance.
	\item It experimentally compares \toolname to the other approach Relda2/RelFix on the same apps used in the latter's evaluation,
	showing that it generally achieves higher precision and recall.
	\item For reproducibility, the implementation of \toolname,
	as well as the details of its experimental evaluation (including the produced fixes), are publicly available in
	a replication package:
	\begin{center}
		\url{https://github.com/bhargavbh/PlumbDROID} \quad (Cite as \cite{PlumbDroid-zenodo})
	\end{center}
\end{itemize}

\section{An Example of \toolname in Action}
\label{sec:example}

IRCCloud is a popular Android app that provides a modern IRC chat client on mobile devices.
\autoref{fig:motivating} shows a (simplified) excerpt of class \J{ImageViewerActivity} in IRCCloud's implementation.

\begin{figure}[!tbh]
\begin{lstlisting}
public class ImageViewerActivity extends Activity {
	
    private MediaPlayer player;
	
    private void onCreate(BundleSavedInstance) {
        // acquire resource MediaPlayer
        player = new MediaPlayer(); (*\label{ln:acquire}*)
        final SurfaceView v = (SufaceView) findViewById(...);
    }
	
    public void onPause() {
        v.setVisibilty(View.INVISIBLE);
        super.onPause();
        (*\label{ln:missingfix}*)// 'player' not released: leak!
    }

}
\end{lstlisting} %
	\caption{An excerpt of class \J{ImageViewerActivity} in Android app IRCCloud,
		showing a resource leak that \toolname can fix automatically.}
	\label{fig:motivating}
\end{figure}

As its name suggests, this class implements the \emph{activity}---a kind of task in Android parlance---triggered
when the user wants to view an image that she downloaded from some chat room.
When the activity starts (method \J{onCreate}),
the class acquires permission to use the system's media player by creating an object of class \J{MediaPlayer} on line~\ref{ln:acquire}.
Other parts of the activity's implementation (not shown here) use \J{player} to interact with the media player
as needed.

When the user performs certain actions---for example, she flips the phone's screen---the Android system
executes the activity's method \J{onPause}, so that the app has a chance to appropriately react to such changes in the environment.
Unfortunately, the implementation of \J{onPause} in \autoref{fig:motivating} does not \emph{release} the media player,
even though the app will be unable to use it while paused \cite{MediaPlayerUsage}.
Instead, it just acquires a new handle to the resource when it resumes.
This causes a \emph{resource leak}: the acquired resource \J{MediaPlayer} is not appropriately released.
Concretely, if the user flips the phone back and forth---thus generating a long sequence of unnecessary new acquires---the leak will result
in wasting system resources and possibly in an overall performance loss.

Such resource leaks can be tricky for programmers to avoid.
Even in this simple example, it is not immediately obvious that \J{onPause} may execute after \J{onCreate},
since this requires a clear picture of Android's reactive control flow.
Furthermore, a developer may incorrectly assume that calling the implementation of \J{onPause} in Android's base class \J{Activity}
(with \J{super.onPause()})
takes care of releasing the held resources.
However, \J{Activity.onPause} cannot know about the resources that have been specifically allocated by the app's implementation;
in particular, it does not release \J{MediaPlayer()} instance \J{player}.

\toolname can automatically analyze the implementation of IRCCloud looking for leaks such as the one highlighted in \autoref{fig:motivating}.
\toolname generates an abstraction of the whole app's control-flow that considers all possible user interactions that may result in leaks.
For each detected leak, \toolname builds a \emph{fix} by adding suitable release statements.

For \autoref{fig:motivating}'s example, \toolname builds a succinct fix at line~\ref{ln:missingfix}
consisting of the conditional release operation \J{if (player != null) player.release()}.
\toolname also checks that the fix is correct (it removes the leak) and ``safe'' (it only releases the resource after the app no longer uses it).
Systematically running \toolname on Android apps can detect and fix many such resource leaks completely automatically. 

\begin{figure*}[!tb]
  \centering
  \begin{adjustwidth}{-20mm}{-5mm}
	\begin{tikzpicture}[node distance=7pt and 10pt,every state/.style={draw,inner sep=0mm,minimum size=6pt}]
	
	\begin{scope}[xshift=0mm]
	\node[inner sep=0pt,label={[above]\textrm{Android app}}] (code) at (0,-15mm)
	{\includegraphics[width=25mm,height=13mm]{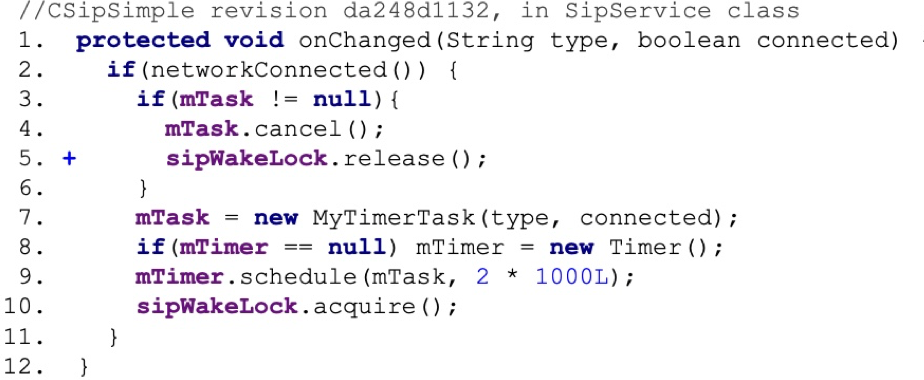}};
	\end{scope}    
	
	\begin{scope}[xshift=30mm]
	\begin{scope}[xshift=0mm,yshift=0mm]
	\draw node[state] (p1-1) {};
	\draw node[state,below right=of p1-1] (p1-2) {};
	\draw node[state,below=of p1-1] (p1-3) {};
	\draw node[state,below=of p1-3] (p1-4) {};
	\begin{scope}[->,every loop/.style={min distance=7pt,out=65,in=5,looseness=5}]
	\path (p1-1) edge (p1-2);
	\path (p1-1) edge (p1-3);
	\path (p1-3) edge (p1-4);
	\path (p1-4) edge (p1-2);
	\path (p1-2) edge (p1-3);
	\end{scope}
	\node[fit=(p1-1)(p1-2)(p1-3)(p1-4),draw,dotted,outer ysep=-2pt,inner xsep=6pt,inner ysep=3pt,label={[above]RFG$_1$}] (rfg1) {};
	\end{scope}
	
	\begin{scope}[xshift=23mm,yshift=-2mm]
	\draw node[state] (p2-1) {};
	\draw node[state,right=of p2-1] (p2-2) {};
	\draw node[state,below=of p2-1] (p2-3) {};
	\draw node[state,below=of p2-2] (p2-4) {};
	\begin{scope}[->,every loop/.style={min distance=7pt,out=65,in=5,looseness=5}]
	\path (p2-1) edge (p2-3);
	\path (p2-3) edge (p2-2);
	\path (p2-3) edge (p2-4);
	\path (p2-2) edge (p2-4);
	\end{scope}
	\node[fit=(p2-1)(p2-2)(p2-3)(p2-4),draw,dotted,outer ysep=-2pt,inner xsep=6pt,inner ysep=5pt,label={[above]RFG$_2$}] (rfg2) {};
	\end{scope}
	
	\begin{scope}[xshift=10mm,yshift=-20mm]
	\draw node[state] (p3-1) {};
	\draw node[state,below=of p3-1] (p3-3) {};
	\draw node[state,below=of p3-3] (p3-4) {};
	\draw node[state,right=of p3-1] (p3-2) {};
	\draw node[state,right=of p3-4] (p3-5) {};
	\begin{scope}[->,every loop/.style={min distance=7pt,out=65,in=5,looseness=5}]
	\path (p3-1) edge (p3-2);
	\path (p3-2) edge (p3-3);
	\path (p3-3) edge (p3-4);
	\path (p3-4) edge (p3-5);
	\path (p3-5) edge (p3-2);
	\path (p3-5) edge (p3-3);
	\end{scope}
	\node[fit=(p3-1)(p3-2)(p3-3)(p3-4)(p3-5),draw,dotted,outer ysep=-2pt,inner xsep=6pt,inner ysep=5pt,label={[above]RFG$_3$}] (rfg3) {};
	\end{scope}
	
	\begin{scope}[->,thick]
	\path (rfg1) edge node[above=-1pt] {\small{}} (rfg2);
	\path ($(rfg1.south)+(0,-2pt)$) edge[bend right] node[sloped,above=0pt] {\small{}} (rfg3.west);
	\path ($(rfg2.south)+(0,-2pt)$) edge[bend left] node[sloped,above=0pt] {\small{}} (rfg3.east);
	\end{scope}
	\node[fit=(rfg1)(rfg2)(rfg3),inner ysep=13pt] (abstraction) {};
	\end{scope}
	
	\begin{scope}[xshift=0mm]
	\node[inner sep=0pt,label={[above]\textrm{Android app}}] (code) at (0,-15mm)
	{\includegraphics[width=25mm,height=13mm]{androidcode2}};
	\end{scope}    
	
	\begin{scope}[xshift=78mm]
	\begin{scope}[xshift=0mm,yshift=0mm]
	\draw node[very thick,vorange,state,label={[below left=-2pt]\color{vorange}$a_k$}] (p1-1) {};
	\draw node[state,below right=of p1-1,opacity=0.3] (p1-2) {};
	\draw node[state,below=of p1-1] (p1-3) {};
	\draw node[state,below=of p1-3] (p1-4) {};
	\begin{scope}[->,every loop/.style={min distance=7pt,out=65,in=5,looseness=5}]
	\path (p1-1) edge[opacity=0.3] (p1-2);
	\path (p1-1) edge[very thick,vorange] (p1-3);
	\path (p1-3) edge[very thick,vorange] (p1-4);
	\path (p1-4) edge[opacity=0.3] (p1-2);
	\path (p1-2) edge[opacity=0.3] (p1-3);
	\end{scope}
	\node[fit=(p1-1)(p1-2)(p1-3)(p1-4),dotted,outer ysep=-2pt,inner xsep=7pt,inner ysep=3pt] (rfg1) {};
	\end{scope}
	
	\begin{scope}[xshift=23mm,yshift=-2mm]
	\draw node[state] (p2-1) {};
	\draw node[state,right=of p2-1] (p2-2) {};
	\draw node[state,below=of p2-1] (p2-3) {};
	\draw node[state,below=of p2-2,opacity=0.3] (p2-4) {};
	\begin{scope}[->,every loop/.style={min distance=7pt,out=65,in=5,looseness=5}]
	\path (p2-1) edge[very thick,vorange] (p2-3);
	\path (p2-3) edge[very thick,vorange] (p2-2);
	\path (p2-3) edge[opacity=0.3] (p2-4);
	\path (p2-2) edge[opacity=0.3] (p2-4);
	\end{scope}
	\node[fit=(p2-1)(p2-2)(p2-3)(p2-4),dotted,outer ysep=-2pt,inner xsep=6pt,inner ysep=5pt] (rfg2) {};
	\end{scope}
	
	\begin{scope}[xshift=10mm,yshift=-20mm]
	\draw node[state] (p3-1) {};
	\draw node[state,below=of p3-1,opacity=0.3] (p3-3) {};
	\draw node[state,below=of p3-3,opacity=0.3] (p3-4) {};
	\draw node[state,right=of p3-1] (p3-2) {};
	\draw node[state,right=of p3-4,opacity=0.3] (p3-5) {};
	\begin{scope}[->,every loop/.style={min distance=7pt,out=65,in=5,looseness=5}]
	\path (p3-1) edge[very thick,vorange] (p3-2);
	\path (p3-2) edge[opacity=0.3] (p3-3);
	\path (p3-3) edge[opacity=0.3] (p3-4);
	\path (p3-4) edge[opacity=0.3] (p3-5);
	\path (p3-5) edge[opacity=0.3] (p3-2);
	\path (p3-5) edge[opacity=0.3] (p3-3);
	\end{scope}
	\node[fit=(p3-1)(p3-2)(p3-3)(p3-4)(p3-5),dotted,outer ysep=-2pt,inner xsep=6pt,inner ysep=5pt] (rfg3) {};
	\end{scope}
	
	\begin{scope}[->,thick]
	\path (rfg1) edge[very thick,vorange] node[above=-1pt] {\small{}} (rfg2);
	\path ($(rfg1.south)+(0,-2pt)$) edge[bend right,opacity=0.3] node[sloped,above=0pt] {\small{}} (rfg3.west);
	\path ($(rfg2.south)+(0,-2pt)$) edge[bend left,very thick,vorange] node[sloped,above=0pt] {\small{}} (rfg3.east);
	\end{scope}
	\node[fit=(rfg1)(rfg2)(rfg3),inner ysep=13pt] (analysis) {};
	\end{scope}
	
	\begin{scope}[xshift=125mm]
	\begin{scope}[xshift=0mm,yshift=0mm]
	\draw node[very thick,vgreen,state,label={[below left=-2pt]\color{vgreen}$a_k$}] (p1-1) {};
	\draw node[state,below right=of p1-1,opacity=0.3] (p1-2) {};
	\draw node[state,below=of p1-1] (p1-3) {};
	\draw node[state,below=of p1-3] (p1-4) {};
	\begin{scope}[->,every loop/.style={min distance=7pt,out=65,in=5,looseness=5}]
	\path (p1-1) edge[opacity=0.3] (p1-2);
	\path (p1-1) edge[very thick,vgreen] (p1-3);
	\path (p1-3) edge[very thick,vgreen] (p1-4);
	\path (p1-4) edge[opacity=0.3] (p1-2);
	\path (p1-2) edge[opacity=0.3] (p1-3);
	\end{scope}
	\node[fit=(p1-1)(p1-2)(p1-3)(p1-4),dotted,outer ysep=-2pt,inner xsep=7pt,inner ysep=3pt] (rfg1) {};
	\end{scope}
	
	\begin{scope}[xshift=23mm,yshift=-2mm]
	\draw node[state] (p2-1) {};
	\draw node[state,right=of p2-1,very thick,vgreen,label={[left=-1pt]\color{vgreen}$r_k$}] (p2-2) {};
	\draw node[state,below=of p2-1] (p2-3) {};
	\draw node[state,below=of p2-2,opacity=0.3] (p2-4) {};
	\begin{scope}[->,every loop/.style={min distance=7pt,out=65,in=5,looseness=5}]
	\path (p2-1) edge[very thick,vgreen] (p2-3);
	\path (p2-3) edge[very thick,vgreen] (p2-2);
	\path (p2-3) edge[opacity=0.3] (p2-4);
	\path (p2-2) edge[opacity=0.3] (p2-4);
	\end{scope}
	\node[fit=(p2-1)(p2-2)(p2-3)(p2-4),dotted,outer ysep=-2pt,inner xsep=6pt,inner ysep=5pt] (rfg2) {};
	\end{scope}
	
	\begin{scope}[xshift=10mm,yshift=-20mm]
	\draw node[state] (p3-1) {};
	\draw node[state,below=of p3-1,opacity=0.3] (p3-3) {};
	\draw node[state,below=of p3-3,opacity=0.3] (p3-4) {};
	\draw node[state,right=of p3-1,very thick,vgreen,label={[above left=-4pt]\color{vgreen}$r_k$}] (p3-2) {};
	\draw node[state,right=of p3-4,opacity=0.3] (p3-5) {};
	\begin{scope}[->,every loop/.style={min distance=7pt,out=65,in=5,looseness=5}]
	\path (p3-1) edge[very thick,vgreen] (p3-2);
	\path (p3-2) edge[opacity=0.3] (p3-3);
	\path (p3-3) edge[opacity=0.3] (p3-4);
	\path (p3-4) edge[opacity=0.3] (p3-5);
	\path (p3-5) edge[opacity=0.3] (p3-2);
	\path (p3-5) edge[opacity=0.3] (p3-3);
	\end{scope}
	\node[fit=(p3-1)(p3-2)(p3-3)(p3-4)(p3-5),dotted,outer ysep=-2pt,inner xsep=6pt,inner ysep=5pt] (rfg3) {};
	\end{scope}
	
	\begin{scope}[->,thick]
	\path (rfg1) edge[very thick,vgreen] node[above=-1pt] {\small{}} (rfg2);
	\path ($(rfg1.south)+(0,-2pt)$) edge[bend right,opacity=0.3] node[sloped,above=0pt] {\small{}} (rfg3.west);
	\path ($(rfg2.south)+(0,-2pt)$) edge[bend left,very thick,vgreen] node[sloped,above=0pt] {\small{}} (rfg3.east);
	\end{scope}
	\node[fit=(rfg1)(rfg2)(rfg3),inner ysep=13pt] (fixing) {};
	\end{scope}
	
	\begin{scope}[vviolet]
	\draw[double,-latex] (code) -- node[above] {\textsc{abstraction}} (abstraction);
	\draw[double,-latex] (abstraction) -- node[above] {\textsc{analysis}} (analysis);
	\draw[double,-latex] (analysis) -- node[above] {\textsc{fixing}} (fixing);
	\draw[double,-latex] (fixing) to[bend left] node[below] {\textsc{validation}} (abstraction);
	\end{scope}
 \end{tikzpicture}
\end{adjustwidth}

	\caption{How \toolname works: 
		First, \toolname builds a finite-state {\color{vviolet}\textsc{abstraction}} of the Android app under analysis,
		which captures \emph{acquire} and \emph{release} operations of an API's resources.
		The abstraction models each function of the application with a \emph{resource-flow graph} (RFG)---a special kind of control-flow graph---and
		combines resource-flow graphs to model inter-procedural behavior.
		In the {\color{vviolet}\textsc{analysis}} step,
		\toolname searches the graph abstraction for {\color{vorange}\emph{resource leaks}}: paths
		where a resource $k$ is acquired ({\color{vorange}$a_k$}) but not eventually released.
		In the {\color{vviolet}\textsc{fixing}} step,
		\toolname injects the missing {\color{vgreen}\emph{release operations $r_k$}}
		where needed along the leaking path.
		In the final {\color{vviolet}\textsc{validation}} step,
		\toolname abstracts and analyzes the code after fixing, so as to ensure that
		the fix does not introduce unintended interactions that cause new resource-usage related problems.
	}
	\label{fig:workflow}
\end{figure*}

\section{How \toolname Works}
\label{sec:how-it-works}

\autoref{fig:workflow} gives a high-level overview of how \toolname works.
Each run of \toolname analyzes an app for leaks of resources from a specific Android API---consisting of acquire and release operations---modeled as described in \autoref{sec:resources}.

The key abstraction used by \toolname is the \emph{resource-flow graph}: a kind of control-flow graph that captures 
the information about possible sequences of acquire and release operations. 
\autoref{sec:rfg} describes how \toolname builds the resource-flow graph for each procedure individually.

A \emph{resource leak} is an execution path where some \emph{acquire} operation is not eventually followed by a matching \emph{release} operation.
In general, absence of leaks (leak freedom) is a \emph{context-free property}~\cite{Sipser}
since there are resources---such as wait locks---that may be acquired and released multiple times (they are \emph{reentrant}).\footnote{For resources that do not allow nesting of acquire and release, leak freedom is a regular property---which \toolname supports as a simpler case.}
Therefore, finding a resource leak is equivalent to analyzing
context-free patterns on the resource-flow graph.
\toolname's detection of resource leaks at the \emph{intra-procedural} level is based on this equivalence, which \autoref{sec:cf-reach}
describes in detail.

\fakepar{Android apps architecture.}
An Android application consists of a collection of standard \emph{components} that have to follow a particular programming model~\cite{AndroidPlatform}. 
Each component type---such as activities, services, and content providers---has an associated \emph{callback graph},
which constrains the order in which user-defined procedures are executed.
As shown by the example of \autoref{fig:CBG},
the \emph{states} of a callback graph are macro-state of the app (such as \emph{Starting}, \emph{Running}, and \emph{Closed}),
connected by edges associated with callback \emph{functions} (such as \J{onStart}, \J{onPause}, and \J{onStop}).
An app's implementation defines procedures that implement the appropriate callback functions of each component (as in the excerpt of \autoref{fig:motivating}).

Following this programming model,
the overall control-flow of an Android app is not explicit from the app's implementation.
Rather, the Android system triggers callbacks according to the transitions that are taken at run time (which, in turn, depend on the events that occur).
\toolname deals with this \emph{implicit} execution flow in two steps.
First (\autoref{sec:explicit-calls}), it defines an \emph{explicit inter-procedural} analysis:
it assumes that the inter-procedural execution order is known,
and combines the intra-procedural analysis of different procedures to
detect leaks across procedure boundaries.
Second (\autoref{sec:callbacks}), it unrolls the callback graph to enumerate sequences of callbacks that may occur when the app is running,
and applies the explicit inter-procedural analysis to these sequences.

\fakepar{Fix generation.}
\toolname's analysis stage extracts detailed information that is useful
not only to detect leaks but also to \emph{generate fixes} that avoid the leaks.
As we will describe in \autoref{sec:fixgen},
\toolname builds fixes by adding a release of every leaked resource as early as possible 
along each leaking execution path.

\toolname's fixes are \emph{correct by construction}:
they release previously acquired resources in a way that guarantees that the previously detected leaks no longer occur.
However, it might still happen that a fix releases a resource that
is used later by the app---thus introducing a use-after-release error. 
In order to rule this out,
\toolname also runs a final \emph{validation step} which reruns the leak analysis on the patched program.
If validation fails, it means that the fix should not be deployed as is; instead, the programmer should modify it in a way that makes it consistent with the rest of the app's behavior.
Our experiments with \toolname (in particular in \autoref{sec:answer-RQ1})
indicate that validation is nearly always successful;
even it fails, it is usually clear how to reconcile the fix with the app's behavior.

\begin{figure}[!tbh]
	\centering
	\begin{tikzpicture}[every state/.style={draw,rectangle,minimum width=14mm,minimum height=10mm}]
	\node[state] (starting) {\emph{Starting}};
	\node[state,below=of starting] (closed) {\emph{Closed}};
	\node[state] (running) at ($(starting)!0.5!(closed)+(47mm,0)$) {\emph{Running}};
	\begin{scope}[-latex]
	\draw (starting) -| node[near start,align=center,fill=white] {\J{onCreate()}\\\J{onStart()}\\\J{onResume()}} (running);
	\draw (running) |- node[near end,align=center,fill=white] {\J{onPause()}\\\J{onStop()}\\\J{onDestroy()}} (closed);
	\draw ($(running.north east)+(-3mm,0)$) -- ++(0,3mm) -- node[right,align=center,fill=white] {\J{onPause()}\\\J{onResume()}} ++(6mm,0) |- (running.east);
	\end{scope}
	
	\end{tikzpicture}
	\caption{Simplified callback graph of an Android component.}
	\label{fig:CBG}
\end{figure}

\subsection{Resources}
\label{sec:resources}
A \toolname analysis targets a specific Android API, which we model as a \emph{resource list} $L$
representing acquire and release 
operations of the API
as a list of pairs $(a_1, r_1)\,(a_2, r_2)\ldots$. 
A pair $(a_k, r_k)$ denotes an operation $a_k$ that \emph{acquires} a certain resource
together with another operation $r_k$ that \emph{releases} the same resource acquired by $a_k$.
The same operation may appear in multiple pairs, corresponding to all legal ways of acquiring and then releasing it.
For simplicity, we sometimes use $L$ to refer to the whole API that $L$ represents.
For example, resource \J{MediaPlayer} can be acquired and released with $(\J{new}, \J{release})$---used in \autoref{fig:motivating}.

\subsection{Intra-Procedural Analysis}
\label{sec:intra-proc}

In the intra-procedural analysis, \toolname
builds a resource-flow graph for every procedure in the app under analysis.
In the example of \autoref{fig:motivating}, it builds one such graph for every callback function, and for all methods called within those functions.

\subsubsection{Resource-Flow Graphs}
\label{sec:rfg}

\toolname's analysis works on a modified kind of control-flow graph called \emph{resource-flow graph} (RFG).
The control-flow graphs of real-world apps are large and complex, but only a small subset of their blocks typically involve accessing resources.
Resource-flow graphs abstract the control flow
by only retaining information that is relevant for detecting resource leaks.\footnote{
	Energy-flow graphs---used by other leak detection techniques for Android~\cite{EnergyPatchTSE18,Relda2TSE16}---are similar to resource-flow graphs
	in that they also abstract control-flow graphs by retaining the essential information
	for leak detection.}

\begin{algorithm}[!htb]
	\KwIn{control-flow basic block $b$, resource list $L$}
	\KwOut{resource path graph $p$}
	\BlankLine
	$p \rec \emptyset$ // initialize $p$ to empty graph \\
	\ForEach{statement $s$ in block $b$}{
		\cIf{$s$ invokes a resource acquire operation $a$ in $L$}{
			$n \rec $ new $\var{AcquireNode}(a)$
		}
		\ucElseIf{$s$ invokes a resource release operation $r$ in $L$}{
			$n \rec$ new $\var{ReleaseNode}(r)$
		}
		\ucElseIf{$s$ invokes any other operation $o$}{
			$n \rec$ new $\var{TransferNode}(o)$
		}
		\ucElseIf{$s$ is a \texttt{\textbf{return}} statement}{
			$n \rec$ new $\var{ExitNode}$
		}
		\cElse{
			$n \rec$ NULL
		}
		\cIf{$n \neq$ NULL}{
			append node $n$ to path graph $p$'s tail
		}
	}
	// if $b$ contains no resource-relevant statements\\
	\cIf{$p = \emptyset$ }{
		$p \rec $ new $\var{TrivialNode}\quad$ // return a trivial node
	}
	\caption{Algorithm \var{Path} that builds the \emph{resource path graph}~$p$ modeling control-flow basic block~$b$.}
	\label{algo:create_rpg}
\end{algorithm}

\begin{algorithm}[!htb]
	\KwIn{control-flow graph $C$, resource list $L$}
	\KwOut{resource-flow graph $R$}
	\BlankLine
	\ForEach{block $b$ in control-flow graph $C$} {
		// $p(b)$ is the path graph corresponding to $b\in C$ \\
		$p(b) \rec \var{Path}(b, L)\quad$  // call to Algorithm~\ref{algo:create_rpg}
	}
	$R \rec \{$ entry node $s\ \}$\\
	$c_0 \rec$ the entry block of $C$ \\
	add an edge from $s$ to $p(c_0)$'s entry\\
	\ForEach{block $b_1$ in control-flow graph $C$}{
		\ForEach{block $b_2$ in $b_1$'s successors in $C$}{
			add an edge from $p(b_1)$'s exit to $p(b_2)$'s entry
		}
		\ForEach{$\var{ExitNode}$ $e$ in $p(b_1)$}{
			add an edge from $e$'s predecessors to $f$ \\(the exit node of $R$)
		} 
	}
	\caption{Algorithm \var{RFG} that builds a \emph{resource-flow graph}~$R$ modeling control-flow graph $C$.}
	\label{algo:cfg2rfg}
\end{algorithm}

A procedure's resource-flow graph $R$ abstracts the procedure's control-flow graph $C$ in two steps.
First, it builds a \emph{resource path} graph $p$ for every \emph{basic block} in $C$---as described by \autoref{algo:create_rpg}.
Then, it builds the resource-flow graph $R$ by connecting the resource path graphs according to the control-flow structure---as described by \autoref{algo:cfg2rfg}.

\fakepar{Resource path graph.}
A basic block corresponds to a sequence of statements without internal branching.
\autoref{algo:create_rpg} builds the resource path graph $p$ for any basic block $b$.
It creates a node $n$ in $p$ for each statement $s$ in $b$ that is relevant to how $L$'s resources are used:
a resource is acquired or released,
or execution terminates with a \J{return}
(which may introduce a leak).
Nodes in the resource path graph also keep track of when any other operation is performed,
because this information is needed for inter-procedural analysis (as we detail in \autoref{sec:inter-proc});
in other words, intra-procedural analysis is sufficient whenever a procedure doesn't have any transfer nodes.
Graph $p$ connects the nodes in the same sequential order as statements in $b$.
When a block $b$ does not include any operations that are relevant for resource usage, its resource path graph $p$ consists of a single \emph{trivial} node, whose only role is to preserve the overall control-flow structure in the resource-flow graph.
Since $b$ is a basic block---that is, it has no branching---$p$ is always a path graph---that is a linear sequence of nodes,
each connected to its unique successor, starting from an entry node and ending in an exit node.

\fakepar{Resource-flow graph.}
\autoref{algo:cfg2rfg} builds the resource-flow graph $R$ of control-flow graph $C$---corresponding to a single procedure.
First, it computes a \emph{path graph} $p(b)$ for every (basic) block $b$ in $C$. 
Then, it connects the various path graphs following the control-flow graph's edge structure:
it initializes $R$ with an entry node $s$ and connects it to the entry node of $p(c_0)$---the path graph of $C$'e entry block;
for every edge $b_1 \to b_2$ connecting block $b_1$ to block $b_2$ in $C$, it connects the exit node of $p(b_1)$ to the entry node of $p(b_2)$.
Since every executable block $b \in C$ is connected to $C$'s entry block $c_0$, and $c_0$'s path graph is connected to $R$'s entry node $s$,
$R$ is a \emph{connected} graph that includes one path subgraph for every executable block in the control-flow graph $C$.
Also, $R$ has a single entry node $s$ and a single exit node $f$.

Given that $R$'s structure matches $C$'s,
if there is a path in $C$ that leaks some of $L$'s resources, there is also a path in $R$ that exposes the same leak, and vice versa.
Thus, we use the expression ``$R$ has leaks/is free from leaks in $L$'' to mean ``the procedure modeled by $C$ has leaks/is free from leaks of resources in the API modeled by $L$''.

\autoref{fig:full-example} shows the (simplified)
resource-flow graphs of methods \J{onCreate} and \J{onPause} from \autoref{fig:motivating}'s example.
Since each method consists of a single basic block, the resource-flow graphs are path graphs (without branching).

\begin{figure*}[!tb]
  \centering
  \begin{adjustwidth}{-22mm}{-5mm}
	\begin{tikzpicture}
	
	\begin{scope}[node distance=5mm and 7mm,
	every state/.style={draw,align=center,rectangle,inner sep=2pt,minimum width=26mm,minimum height=7mm}]
	\begin{scope}[xshift=0mm,yshift=0mm]
	\draw node[state,label=left:\textsl{AcquireNode}] (os-1) {\color{vorange}$\Jm{new}$};
	\draw node[state,below=of os-1,label=left:\textsl{TransferNode}] (os-2) {$\Jm{findViewById()}$};
	\end{scope}
	\begin{scope}[xshift=50mm,yshift=0mm]
	\draw node[state,label=right:\textsl{TransferNode}] (op-1) {$\Jm{setVisibility()}$};
	\draw node[state,below=of op-1,label=right:\textsl{TransferNode}] (op-2) {$\Jm{super.onPause()}$};
	\end{scope}
	\draw node[state,right=33mm of op-2,label=right:\textsl{ReleaseNode}] (op-fix) {\color{vgreen}$\Jm{player.release()}$};
	\end{scope}
	
	\begin{scope}[->,very thick,vorange]
	\path (os-1) edge (os-2);
	\path (op-1) edge (op-2);
	\end{scope}
	
	\begin{scope}[node distance=7mm and 5mm,,
	every state/.style={draw,inner sep=2pt,minimum size=8mm}]
	\draw node[state,above=of os-1] (os-s) {$s$};
	\draw node[state,below=of os-2] (os-f) {$f$};
	\draw node[state,above=of op-1] (op-s) {$s$};
	\draw node[state,below=of op-2] (op-f) {$f$};
	\end{scope}
	
	\node[fit=(os-s)(os-1)(os-2)(os-f),draw,dotted,thick,outer ysep=-2pt,inner xsep=3pt,inner ysep=7pt,label={\J{onCreate}'s RFG}] (rfg-os) {};
	\node[fit=(op-s)(op-1)(op-2)(op-f),draw,dotted,thick,outer ysep=-2pt,inner xsep=3pt,inner ysep=7pt,label={\J{onPause}'s RFG}] (rfg-op) {};

	\begin{scope}[->,very thick,vorange]
	\path (os-s) edge (os-1);
	\path (os-2) edge (os-f);
	\path (op-s) edge (op-1);
	\path (op-2) edge (op-f);
	\end{scope}
	
	\begin{scope}[->,very thick,vorange]
	\draw let
	\p1 = ($(os-s)!0.5!(op-s)$),
	\p2 = (os-s),
	\p3 = (os-f)
	in
	(os-f) -- (\x1,\y3) -- (\x1,\y2) -- (op-s);
	\end{scope}
	
	\begin{scope}[->,very thick,vgreen]
	\draw let
	\p1 = (op-2.east),
	\p2 = (op-fix.west),
	\p3 = ($(\p1)+(0,-6pt)$)
	in
	(\p3) -- (\x2,\y3);
	\draw (op-fix.south) -- (op-f);
	\end{scope}
	
	\node at ($(op-fix.north)+ (7mm,7mm)$) {\emph{\toolname's fix introduces node:}};
	
 \end{tikzpicture}
\end{adjustwidth}
\caption{Some abstractions built by \toolname to analyze the example of \autoref{fig:motivating}. 
		The intra-procedural analysis described in \autoref{sec:intra-proc} builds a resource-flow graph (RFG) for each procedure \J{onCreate}
		and \J{onPause} independently.
		As described in \autoref{sec:inter-proc},
		the inter-procedural analysis considers, among others, the sequence of callback functions \J{onCreate(); onStart(); onResume(); onPause()};
		since \J{onStart()} and \J{onResume()} do nothing in this example, this corresponds to connecting \J{onCreate}'s exit to
		\J{onPause}'s entry.
		The inter-procedural analysis thus finds a \emph{leaking path} (in {\color{vorange}orange}),
		where acquire operation \J{new} is not matched by any release operation.
		Fixing (\autoref{sec:fixgen}) modifies the app by adding a suitable release operation \J{player.release()} (in {\color{vgreen}green}),
		which completely removes the leak.
	}
	\label{fig:full-example}
\end{figure*}

\subsubsection{Context-Free Emptiness: Overview}
\label{sec:cf-reach}

Given a resource-flow graph $R$---abstracting a procedure $P$ of the app under anal\-y\-sis---and a resource list $L$,
$P$ is free from leaks of resources in $L$ if and only if every execution trace in $R$ consistently acquires and releases resources in $L$.
We express this check as a formal-language inclusion problem---\`a~la automata-based model-checking~\cite{VardiW86}---as follows (see \autoref{fig:abmc} for a graphical illustration):

\begin{figure}[!tb]
	\centering
	\begin{tikzpicture}
	\draw[fill,color=vorange,opacity=0.3] (0,0) rectangle +(35mm, 30mm);
	\draw[fill,color=vgreen,opacity=0.3] (35mm,0) rectangle +(45mm, 30mm);
	\node[color=vgreen,fill=white] (AL) at (68mm,3.5mm) {$A_L$: leak-free};
	\node[left=35mm of AL,color=vorange,fill=white] (cAL) {$\overline{A_L}$: leaking};
	\draw[fill,color=vviolet,opacity=0.3] (10mm,10mm) rectangle +(60mm,18mm);
	\node[color=vviolet,fill=white] (AR) at (41.5mm,23mm) {$A_R$: procedure's resource usages};
	\node[align=center] (AX) at (23mm,15mm) {$A^X$: \\[-2pt]leaking usages};
	\draw[ultra thick,dotted,color=white] (10mm,10mm) rectangle +(25mm,18mm);
	\end{tikzpicture}
	
	\caption{A graphical display of \toolname's intra-procedural leak detection.
		The \emph{resource} automaton {\color{vgreen} $A_L$} captures all
		{\color{vgreen} leak-free sequences}; its \emph{complement} {\color{vorange} $\overline{A_L}$} captures all {\color{vorange} leaking sequences}.
		The \emph{flow} automaton {\color{vviolet} $A_R$} captures all sequences of
		{\color{vviolet} resource usages} that may happen when the procedure under analysis runs.
		The \emph{intersection} automaton $A^X$ captures
		the intersection of
		the {\color{vviolet} violet} and {\color{vorange} orange} areas
		(outlined in dotted white), which
		marks the procedure's resource usage sequences that are leaking.
		If this area is empty, we conclude that the procedure is leak free.
	}
	\label{fig:abmc}
\end{figure}

\begin{enumerate}
	\item We define a \emph{resource automaton} $A_L$ that accepts sequences of operations
	in $L$ that are free from leaks.
	Since leak freedom is a (deterministic)
	context-free property,
	the resource automaton is a (deterministic) pushdown automaton.\footnote{We could equivalently use context-free grammars.}
	
	(Resource automata are described in \autoref{step:resource-automata}.)
	
	\item We define the \emph{complement} automaton $\overline{A_L}$ of $A_L$,
	which accepts precisely all sequences of operations in $L$ that \emph{leak}.
	Since $A_L$ is a deterministic push-down automaton, $\overline{A_L}$ is
	too a deterministic push-down automaton.
	
	(Complement automata are described in \autoref{step:complement}.)
	
	\item We define a \emph{flow automaton} $A_R$ that captures all possible paths
	through the nodes of resource-flow graph $R$.
	Since $A_R$ is built directly from $R$, it is a (deterministic) finite-state automaton.
	
	(Flow automata are described in \autoref{step:flow}.)
	
	\item We define the intersection automaton $A^X = \overline{A_L} \times A_R$ that accepts
	all possible paths in $R$ that introduce a leak in resource $L$.
	
	(Intersection automata are described in \autoref{step:intersection}.)
	
	\item We check if the intersection automaton accepts no inputs (the ``empty language'').
	If this is the case, it means that $R$ cannot leak; otherwise, we found a leaking trace.
	
	(Emptiness checking is described in \autoref{step:emptiness}.)
\end{enumerate}

The following subsections present these steps in detail.

\begin{figure*}[!tb]
	\begin{subfigure}[b]{0.45\textwidth}
		\centering
		\begin{tikzpicture}[node distance=10mm and 20mm]
		\node[state,initial,initial distance=3mm,initial above,initial text={}] (q0) {};
		\node[state,right=of q0] (q1) {};
		\node[state,right=of q1,accepting] (q2) {};
		\begin{scope}[->,every loop/.style={min distance=7pt,looseness=5}]]
		\path (q0) edge node[above] {$s, \bot \colon \bot$} (q1);
		\path (q1) edge node[above] {$f, \bot\colon \bot$} (q2);
		\path (q1) edge[loop above] node[above] {$\Jm{acquire}, * \colon \Jm{acquire}\,*$} ();
		\path (q1) edge[loop below] node[below] {$\Jm{release}, \Jm{acquire} \colon \epsilon$} ();
		\end{scope}
		\end{tikzpicture}
		\caption{Deterministic pushdown automaton $A_{L_W}$ accepting the language of leak-free sequences of \J{acquire} and \J{release} operations of resource \J{WifiLock}.
			A transition $x, y \colon Z$ reads input $x$ when $y$ is on top of the stack, and replaces $y$ with string $Z$.
			$*$ is a shorthand for any symbol, $\epsilon$ is the empty string, and $\bot$ is the empty stack symbol.
			The short arrow in the leftmost state denotes that it is an initial state;
			the double line in the rightmost state denotes that it is a final state.} 
		\label{fig:pda-nonleak-wifilock}
	\end{subfigure}
	\hfill
	\begin{subfigure}[b]{0.45\textwidth}
		\centering
		\begin{tikzpicture}
		\matrix [row sep=7mm, column sep=20mm] {
			& \node[state,initial,initial distance=3mm,initial above,initial text={}] (q0) {}; &  \\
			\node[state] (q2) {}; & \node[state] (q1) {}; & \node[state] (q3) {}; \\
			& \node[state,accepting] (q4) {}; &  \\
		};
		\begin{scope}[->,every loop/.style={min distance=7pt,looseness=5}]]
		\path (q0) edge node[left] {$s$} (q1);
		\path (q1) edge node[left] {$f$} (q4);
		\path (q1) edge[bend right] node[above] {\J{new}} (q2);
		\path (q2) edge[bend right] node[below] {\J{release}} (q1);
		\path (q1) edge[bend left] node[above] {\J{start}} (q3);
		\path (q3) edge[bend left] node[below] {\J{stop}} (q1);
		\end{scope}
		\end{tikzpicture}
		\caption{Deterministic finite-state automaton $A_{L_M}$ accepting the language of leak-free sequences of operations of resource \J{MediaPlayer}.
			The short arrow in the top state denotes that it is an initial state;
			the double line in the bottom state denotes that it is a final state.} 
		\label{fig:fsa-nonleak-mediaplayer}
	\end{subfigure}
	\caption{Resource automata for reentrant resource \J{WifiLock} and non-reentrant resource \J{MediaPlayer}.}
	\label{fig:two-res-automata}
\end{figure*}

\begin{figure}[!bt]
	\centering
	\begin{tikzpicture}
	
	\begin{scope}[node distance=5mm and 3mm,
	every state/.style={draw,align=center,inner sep=0pt,minimum size=12pt}]
	
	\begin{scope}[xshift=25mm,yshift=14mm]
	\node[state,initial,initial distance=3mm,initial above,initial text={}] (f0) {$s$};
	\node[state,below=of f0] (f1) {};
	\node[state,below=of f1] (f2) {};
	\node[state,below=of f2] (f3) {$f$};
	\node[state,below=of f3,accepting] (f4) {$e$};
	\end{scope}
	
	\begin{scope}[->]
	\path (f0) edge node[left] {$s$} (f1);
	\path (f1) edge node[left] {\J{new}} (f2);
	\path (f2) edge (f3);
	\path (f3) edge node[left] {$f$} (f4);
	\end{scope}
	
	\matrix [row sep=10mm, column sep=7mm] {
		& \node[state,initial,initial distance=3mm,initial above,initial text={}] (q0) {}; &  \\
		\node[state] (q2) {}; & \node[state] (q1) {}; & \node[state] (q3) {}; \\
		& \node[state,accepting] (q4) {}; &  \\
	};
	\begin{scope}[->,every loop/.style={min distance=7pt,looseness=5}]]
	\path (q0) edge node[left,near start] {$s$} (q1);
	\path (q1) edge[bend right] node[above] {\J{new}} (q2);
	\path (q2) edge[bend right] node[below] {\J{release}} (q1);
	\path (q1) edge[bend left] node[above] {\J{start}} (q3);
	\path (q3) edge[bend left] node[below] {\J{stop}} (q1);
	\path (q2) edge[out=-100,in=180] node[below,near end] {$f$} (q4);
	\path (q3) edge[out=-80,in=0] node[below,near end] {$f$} (q4);
	\end{scope}
	
	\begin{scope}[xshift=48mm,yshift=14mm]
	\node[state,initial,initial distance=3mm,initial above,initial text={}] (i0) {$s$};
	\node[state,below=of i0] (i1) {};
	\node[state,below=of i1] (i2) {};
	\node[state,below=of i2] (i3) {$f$};
	\node[state,below=of i3,accepting] (i4) {$e$};
	
	\begin{scope}[->]
	\path (i0) edge node[left] {$s$} (i1);
	\path (i1) edge node[left] {\J{new}} (i2);
	\path (i2) edge (i3);
	\path (i3) edge node[left] {$f$} (i4);
	\end{scope}
	\end{scope}
	
	\end{scope}
	
	\node[above=5mm of f0] (AR) {$A_R$};
	\path let \p1 = (AR), \p2 = (q0) in node at (\x2,\y1) {$\overline{A_{L_M}}$};
	\path let \p1 = (AR), \p2 = (i0) in node at (\x2,\y1) {$A^X = \overline{A_{L_M}} \times A_R$};
	\end{tikzpicture}
	
	\caption{The intersection automaton $A^X$ (right) combines
		the complement automaton $\overline{A_{L_M}}$ (left)
		of the \J{MediaPlayer}'s resource automaton $A_{L_M}$ from \autoref{fig:fsa-nonleak-mediaplayer}
		and the flow automaton $A_R$ (middle) of method \J{onCreate()} from \autoref{fig:motivating}.
	}
	\label{fig:automata-examples}
\end{figure}

\subsubsection{Resource Automata}
\label{step:resource-automata}

Given a resource list $L = \{ (a_1, r_1)\, (a_2, r_2)\, \ldots\, (a_n, r_n) \}$,
the \emph{resource automaton} $A_L$ is 
a deterministic pushdown automaton\footnote{Precisely, a visibly pushdown automaton (a subclass of deterministic pushdown automata~\cite{EmptinessPDA})
	would be sufficient.}
that accepts all strings
that begin with a start character $s$,
end with a final character $f$, and
include all sequences of the characters $a_1, r_1, a_2, r_2, \ldots, a_n, r_n$
that encode all leak-free sequences of acquire and release operations.

More precisely, $A_L$ is a deterministic pushdown automaton if the resource modeled by $L$ is a \emph{reentrant},
and hence it can be acquired multiple times.
If the resource modeled by $L$ is not reentrant, $A_L$ is an ordinary finite state automaton (a simpler subclass of pushdown automata).
In the remainder of this section,
we recall the definitions of deterministic pushdown automata and finite-state automata,
and we show the example of resource automaton of a reentrant resource, as well as one of a non-reentrant resource.

\fakepar{Pushdown automata.}
Pushdown automata~\cite{Sipser} are finite-state automata equipped with
an unbounded memory that is manipulated as a stack.
For leak detection, we only need \emph{deterministic} pushdown automata,
where the input symbol uniquely determines the transition to be taken in each state.

\begin{definition}[Deterministic pushdown automaton]
	A \emph{deterministic pushdown automaton} $A$ is a tuple $\langle \Sigma, Q, I, \Gamma, \delta, F \rangle$,
	where:
	\begin{enumerate*}
		\item $\Sigma$ is the input alphabet;
		\item $Q$ is the set of control states;
		\item $I \subseteq Q$ and $F \subseteq Q$ are the sets of initial and final states;
		\item $\Gamma$ is the stack alphabet, which includes a special ``empty stack'' symbol $\bot$;
		\item and $\delta \colon Q \times \Sigma \times \Gamma \to Q \times \Gamma^*$ is the transition function.
	\end{enumerate*}
	An automaton's computation starts in an initial state with an empty stack $\bot$.
	When the automaton is in state $q_1$ with stack top symbol $\gamma$ and input $\sigma$,
	if $\delta(q_1, \sigma, \gamma) = (q_2, G)$ is defined,
	it moves to state $q_2$ and replaces symbol $\gamma$ on the stack with string $G$.
	$\lang(A) \subseteq \Sigma^*$ denotes the set of all input strings $s$ accepted by $A$, that is such that $A$ can go from one of its initial states to one of its final states by inputting $s$.
\end{definition}

\fakepar{Reentrant resources.}
Consider a \emph{reentrant} resource with resource list \linebreak
$L = \{ (a_1, r_1)\, (a_2, r_2)\, \ldots\, (a_n, r_n) \}$.  The
resource automaton $A_L$ is a deterministic pushdown automaton that
operates as follows.  When $A^L$ inputs an acquire operation $a_k$,
it pushes it on to the stack; when it inputs a release operation
$r_k$, if the symbol on top of the stack is some $a_k$ such that
$(a_k, r_k) \in L$, it pops $a_k$---meaning that the release matches
the most recent acquire: since all operations in $L$ correspond to
the same resource, any release has to refer to the latest acquire.
Finally, $A^L$ accepts the input if it ends up with an empty stack.

\begin{example}[Resource automaton for \texttt{WifiLock}]
	Let's illustrate this construction in detail for the case of resource \J{WifiLock}
	with $L_W = \{ (\Jm{acquire}, \Jm{release}) \}$ that is reentrant (see \autoref{tab:resources}).
	Pushdown automaton $A_{L_W}$ in \autoref{fig:pda-nonleak-wifilock}
	accepts all strings over alphabet $\Sigma^{L_W} = \{ s, f, \Jm{acquire}, \Jm{release}  \}$ of the form $s\,B\,f$
	where $B$ is a string $B \in \{\Jm{acquire}, \Jm{release}\}^*$ is any balanced sequence of \J{acquire} and \J{release}---that is, a leak-free sequence.
\end{example}

\fakepar{Finite-state automata.}
Finite-state automata can be seen as a special case of pushdown automata
without stack.

\begin{definition}[Finite-state automaton]
	A \emph{finite-state automaton} $A$ is a five-element tuple $\langle \Sigma, Q, I, \delta, F \rangle$,
	where:
	\begin{enumerate*}
		\item $\Sigma$ is the input alphabet;
		\item $Q$ is the set of control states;
		\item $I \subseteq Q$ and $F \subseteq Q$ are the sets of initial and final states;
		\item and $\delta \subseteq Q \times \Sigma \times Q$ is the transition relation. 	\end{enumerate*}
	An automaton's computation starts in an initial state.
	When the automaton is in state $q_1$ with input $\sigma$,
	if $q_2 \in \delta(q_1, \sigma)$,
	it may move to state $q_2$.
	$\lang(A) \subseteq \Sigma^*$ denotes the set of all input strings $s$ accepted by $A$, that is such that $A$ can go from one of its initial states to one of its final states by inputting $s$.
\end{definition}

A finite-state automaton is \emph{deterministic} when its transition relation $\delta$ is
actually a function: the input uniquely determines the next state.

\fakepar{Non-reentrant resources.}
Consider a \emph{non}-reentrant resource with resource list
$L = \{ (a_1, r_1)\, (a_2, r_2)\, \ldots\, (a_n, r_n) \}$.  The
resource automaton $A_L$ is a finite-state automaton that
operates as follows.  When $A^L$ inputs an acquire operation $a_k$,
it moves to a new fresh state that is not final;
the only valid transitions out of this state correspond to release operations
$r_k$ such that $(a_k, r_k) \in L$.
Thus, $A^L$ accepts the input only if every acquire operation is immediately followed by a matching release operation.

\begin{example}[Resource automaton for \texttt{MediaPlayer}]
	Resource \J{MediaPlayer} (used in the example of \autoref{sec:example})
	is instead not reentrant (see \autoref{tab:resources}),
	and offers operations $L_{M} = \{ (\Jm{new}, \Jm{release})\, (\Jm{start}, \Jm{stop}) \}$.
	The finite-state automaton $A_{L_M}$ in \autoref{fig:fsa-nonleak-mediaplayer}
	accepts all strings over alphabet $\Sigma^{L_M} = \{ s, f, \Jm{new}, \Jm{release}, \Jm{start}, \Jm{stop} \}$ of the form $s\,(\Jm{new}\,\Jm{release}\mid\Jm{start}\,\Jm{stop})^*\,f$,
	where each acquire operation is immediately followed by the matching release operation---that is,
	all leak-free sequences.
\end{example}

\subsubsection{Complement Automata}
\label{step:complement}

Deterministic pushdown automata are a strict subclass of (nondeterministic) pushdown automata~\cite{EmptinessPDA}.
Unlike general pushdown automata, deterministic pushdown automata
are closed under \emph{complement}.
That is, given any deterministic pushdown automaton $A$,
we can always build the complement $\overline{A}$: a deterministic pushdown automaton
that accepts precisely the inputs that $A$ rejects and vice versa.
For brevity, we do not repeat the classic construction of complement automata~\cite{Sipser}
The key idea is to switch final and non-final states, so
that every computation that ends in a final state in $A$ will be rejected in $\overline{A}$ and vice versa.
Since finite-state automata are a subclass of deterministic pushdown automata, they are closed under complement too.

For our purposes, we need a slightly different complement automaton: one that accepts all sequences that begin with $s$, end with $f$, and
include any leaking sequence of acquire and release in between these markers.
For example, the automaton on the left in \autoref{fig:automata-examples} is the complement of
\J{MediaPlayer}'s resource automaton in \autoref{fig:fsa-nonleak-mediaplayer} that \toolname builds:
it only accepts sequences that terminate with an $f$ when there is a pending acquired resource before it's released.

\subsubsection{Flow Automata}
\label{step:flow}
Given a resource-flow graph $R = \langle V, E \rangle$,
the \emph{flow automaton} $A_R$
is a deterministic finite-state automaton that accepts precisely the language $\lang(R)$
of all paths $\pi$ through $R$ such that $\pi$ starts in $R$'s entry node $s$ and ends in $R$'s exit node~$f$.

More precisely, the flow automaton $A_R$
is a tuple $\langle \Sigma^R, Q^R, I^R, \delta^R, F^R \rangle$,
where:
\begin{enumerate*}
	\item $\Sigma^R = \Sigma^L$ are all acquire and release operations, plus symbols $s$ and $f$;
	\item $Q^R = V \cup \linebreak \{e\}$ are all nodes of $R$ plus a fresh exit node $e$;
	\item $I^R = \{ s \}$ is the unique entry node of $R$, and $F^R = \{ e \}$ is the new unique exit node;
	\item the transition relation $\delta^R$ derives from $R$'s edges:
	for every edge $m \to n$ in $R$, $n \in \delta^R(m, \var{type}(m))$
	is a transition from state $m$ to state $n$ that reads input symbol $\var{type}(m)$
	corresponding to the type of node $m$ (start, acquire, or release);
	plus a transition from $R$'s exit node to the new exit node $e$ reading $f$.
\end{enumerate*}

Without loss of generality, we can assume that $A_R$ is \emph{deterministic},
since finite-state automata are closed under determinization~\cite{Sipser}.
That is, even if the construction above gives a nondeterministic finite-state automaton $A_R$,
we can always build an equivalent deterministic variant of it that accepts precisely the same inputs.

The middle automaton in \autoref{fig:automata-examples}
is the flow automaton $A_R$ of \J{onCreate} in \autoref{fig:motivating}'s running example:
it is isomorphic to \J{onCreate}'s resource-flow graph (left in~\autoref{fig:full-example})
except for an additional final node $e$ that follows $f$.

\subsubsection{Intersection Automata}
\label{step:intersection}

Given (the complement of) a resource automaton $\overline{A_L}$ and flow automaton $A_R$,
the \emph{intersection} automaton $A^X = \overline{A_L} \times A_R$
is a deterministic pushdown automaton that accepts
precisely the intersection language $\overline{\lang(A_L)} \cap \lang(A_R)$,
that is the inputs accepted by both the complement automaton $\overline{A_L}$
and the flow automaton $A_R$.
Therefore, $A^X$ precisely captures all sequences of acquire and release operations
that may occur in $R$ (that is, they are in $\lang(A_R)$)
and that leak some resource in $L$ (that is, they are in $\overline{\lang(A_L)} = \lang(\overline{A_L})$ and thus are rejected by $A_L$).

More precisely, the intersection automaton $A^X$ is a deterministic pushdown automaton
$\langle \Sigma^X, Q^X, I^X, \Gamma^X, \delta^X, F^X \rangle$, where:
\begin{enumerate*}
	\item $\Sigma^X = \Sigma^L$ is the usual alphabet of all acquire and release operations, plus symbols $s$ and $f$;
	\item $Q^X = Q^R \times Q^L$ is the Cartesian product of $A_R$'s and $\overline{A_L}$'s states;
	\item $I^X = (i_1, i_2)$, where $i_1 \in I^R$ is an initial state of $A_R$ and $i_2 \in I^L$ is an initial state of $\overline{A_L}$;
	\item $F^X = (f_1, f_2)$, where $f_1 \in F^R$ is a final state of $A_R$ and $f_2 \in F^L$ is a final state of $\overline{A_L}$;
	\item $\Gamma^X = \Gamma^L$ is the stack alphabet of $\overline{A_L}$;
	\item for every transition $p_2 \in \delta^R(p_1, \sigma)$ in $A_R$
	and every transition $(q_2, G) = \delta^L(q_1, \sigma, \gamma)$ in $\overline{A_L}$
	that input the same symbol $\sigma$,
	$A^X$ includes transition $((p_2, q_2), G) = \delta^X((p_1, q_1), \sigma, \gamma)$
	that manipulates the stack as in $\overline{A_L}$'s transition.
\end{enumerate*}
Since both $A_R$ and $\overline{A_L}$ are deterministic, so is $A^X$.

The rightmost automaton in \autoref{fig:automata-examples}
is the intersection automaton of \autoref{fig:motivating}'s running example.
Since the flow automaton $A_R$ is just a single path,
it is identical to its intersection with $\overline{A_{L_M}}$, which accepts the single leaking path.

\subsubsection{Emptiness Checking}
\label{step:emptiness}

In the last step of its intra-procedural leak detection,
\toolname has built the intersection automaton $A^X$:
a deterministic pushdown automaton that accepts
all sequences of operations on resources $L$
that may occur in the piece of code $P$ corresponding to $R$ and
that \emph{leak} some resource.
In other words, $R$ is free from leaks in $L$ if and only if
the intersection automaton $A^X$ accepts the \emph{empty} language---that is, no inputs at all.

It is another classic result of automata theory
that checking whether any pushdown automaton accepts the empty language
(that is, it accepts no inputs) is decidable in polynomial time
\cite{EmptinessPDA}.

\toolname applies this classic decision procedure
to determine whether $A^X$ accepts the empty language.
If it does, then procedure $P$ is leak free;
otherwise, we found a \emph{trace} of $P$ that leaks.

\subsection{Inter-Procedural Analysis}
\label{sec:inter-proc}

\toolname lifts the intra-procedural analysis to a whole app by analyzing all possible calls between procedures.
The analysis of a given sequence of procedure calls combines the results of intra-procedural analysis as described in \autoref{sec:explicit-calls}.
Since in Android system callbacks determine the overall execution order of an app,
\autoref{sec:callbacks} explains how \toolname unrolls the callback graph to
enumerate possible sequences of procedure calls---which are analyzed as if they were an explicit call sequence.

\begin{algorithm}[!htb]
	\KwIn{call graph $C = \langle V, E \rangle$, automaton $\overline{A^L}$} 
	\KwOut{ $H = \left\{H_p \mid V \ni p \text{ is not called by any procedure} \right\}$}
	\BlankLine
	$N \rec \text{ topological sort of }C$ \label{ln:inter:toposort}\\
	\ForEach{$n \in N$}{ // for each procedure $n$ \\
		\ucIf{$n$ is not calling any other procedure \label{ln:inter:bottom}}{ 
			// leaking paths in \\
			// intra-procedural analysis of $n$ \\
			// $R_n$ is the RFG of procedure $n$\\
			$H_n \rec \var{LeakingPaths}(R_n, \overline{A^L}, L)$ }
		\ucElseIf { $n$ calls procedures $m_1,m_2,\ldots$ \label{ln:inter:caller}} {
			$R_n' \rec R_n$ \\
			\ForEach{ $m \in \{ m_1,m_2,\ldots \}$} {
				// $R_n'$ is $R_n$ with call-to-$m$ nodes \\
				// replaced by $H_m$ \\
				$R_n' \rec R_n'[\var{TransferNode}(m) \mapsto H_m]$
			}// leaking paths in \\
			// intra-procedural analysis of $R_n'$ \\
			$H_n \rec \var{LeakingPaths}(R_n', \overline{A^L}, L)$ \label{ln:inter:endcaller}
		}}
		\caption{Algorithm $\var{AllCalls}$ which computes inter-procedural resource-flow paths accepted by ``leaking'' pushdown automaton $\overline{A^L}$. Function $\var{LeakingPaths}$ performs the intra-procedural detection of leaking paths described in \autoref{sec:cf-reach}.}
		\label{algo:interprocedural}
	\end{algorithm}
	
	\subsubsection{Explicit Call Sequences}
	\label{sec:explicit-calls}
	
	As it is customary, \toolname models calls between procedures with a \emph{call graph} $C$:
	every node $v$ in $C$ is one of the procedures that make the app under analysis;
	and an edge $u \to v$ in $C$ means that $u$ calls $v$ directly.
	In our analysis, a call graph may have multiple entry nodes, since Android applications have multiple entry points.
	
	\toolname follows \autoref{algo:interprocedural} to perform inter-procedural analysis based on the call graph.
	First of all, we use topological sort (line~\ref{ln:inter:toposort}) to rank $C$'s nodes in an order that is consistent with the call order encoded by $C$'s edges:
	if a node $P$ has lower rank than a node $Q$ it means that $P$ does not call $Q$.
	Topological sort is applicable only if $C$ is acyclic, that is there are no circular calls between procedures.
	If it detects a cycle, \toolname's implementation issues a warning and then breaks the cycle somewhere.
	As we discuss in \autoref{sec:limitations} and \autoref{sec:experiments},
	the restriction to acyclic call graphs seems minor in practice since all apps we analyzed had acyclic call graphs.
	
	Once nodes in $C$ are ranked according to their call dependencies,
	\autoref{algo:interprocedural} processes each of them starting from those corresponding to procedures that do not call any other procedures (line~\ref{ln:inter:bottom}).
	The resource-flow graph of such procedures doesn't have any \emph{transfer} nodes, and hence it can be completely analyzed using
	intra-procedural analysis.
	
	Function $\var{LeakingPaths}$ performs the intra-procedural leak
	detection technique of \autoref{sec:cf-reach}
	and returns any \emph{leaking paths} in the procedure.
	The leaking path, if it exists, is used as a \emph{summary} of the procedure.
	
	Procedures that are free from leaks have an empty path as summary;
	therefore, they are neutral for inter-procedural analysis.
	In contrast, procedures that may leak have some non-empty path as summary,
	which can be combined with the summary of other procedures they call to find out whether the combination of caller and callee is free from leaks.
	This is done in lines~\ref{ln:inter:caller}--\ref{ln:inter:endcaller} of \autoref{algo:interprocedural}:
	the resource-flow graph of a procedure $n$ that calls another procedure $m$ includes some transfer nodes to $m$;
	we replace those nodes with the summary of $m$ (which was computed before thanks to the topological sorting),
	and perform an analysis of the call-free resource-flow graph with summaries.
	The output of \autoref{algo:interprocedural} are complete summaries for the whole app starting from the entry points.
	
	\subsubsection{Implicit Call Sequences}
	\label{sec:callbacks}
	Callbacks in every component used by an Android app have to follow an execution order given by the component's callback graph:
	a finite-state diagram with callback functions defined on the edges (see \autoref{fig:CBG} for a simplified example).
	Apps provide implementations of such callback functions, which \toolname can analyze for leaks.
	When an edge's transition is taken, \emph{all} callback functions defined on the edge are called in the order in which they appear.
	
	The documentation of every resource defines callback functions 
	where the resource ought to be released.
	\toolname enumerates all paths that first go from the callback graph's entry to the nodes from where the release callback functions can be called,
	and then continue looping until states are traversed up to $D$ times---where $D$ is a configurable parameter of \toolname called ``unrolling depth'' (see \autoref{sec:answer-RQ5} to see its impact in practice).
	Each unrolled path determines a sequence of procedures $P_1; P_2; \ldots$ used in the callback functions in that order.
	\toolname looks for leaks in these call sequences by analyzing them as if they were explicit calls in that sequence---using the approach of \autoref{sec:explicit-calls}.
	
	For example, \autoref{fig:motivating}'s resource \J{MediaPlayer} should be released in callback function \J{onPause}.
	For a component with the callback graph of \autoref{fig:CBG}, and unrolling depth $D = 2$,
	\toolname enumerates the path $\var{Starting} \to \var{Running} \to \var{Running} \to \var{Closed}$, corresponding to callback sequence
   \J{onCreate();} \J{onStart();} \J{onResume();} \J{onPause();} \J{onResume();} \J{onPause();} \ldots.
	If the media player is acquired and not later released in these call's implementations, \toolname will detect a leak.
	Since callback functions \J{onStart} and \J{onResume}  are not implemented in \autoref{fig:motivating}'s simple example,
	\autoref{fig:full-example} displays the initial part of this sequence of callbacks by connecting in a sequence
	the resource-flow graphs of \J{onCreate} and \J{onPause}.

	\subsection{Fix Generation}
	\label{sec:fixgen}
	
	\fakepar{Fix templates.}
	Once \toolname detects a resource leak, fixing it amounts to injecting missing release operations at suitable locations in the app's implementation.
	\toolname builds fixes using the conditional template: \\$\J{if (resource != null && }\ \;\var{held}\J{)}\:\J{resource.}r\J{()}$, where \J{resource} is a reference to the resource object, $r$ is the release operation (defined in the resource's API), and \var{held} is a condition that holds if and only if the resource is actually not yet released.
	Calls to release operations must be conditional because \toolname's analysis is an over-approximation (see \autoref{sec:limitations}) and, in particular, a \emph{may leak} analysis~\cite{ProgramAnalysisBook}:
	it is possible that a leak occurs only in certain conditions, but the fix must be correct in all conditions.
	Condition \var{held} depends on the resource's API:
	for example, wake locks have a method \J{isHeld()} that perfectly serves this purpose;
	in other cases, the null check is enough (and hence \var{held} is just \J{true}).
	Therefore, \toolname includes a definition of \var{held} for every resource type, which it uses to instantiate the template.
	
	Another complication in building a fix arises when a reference to the resource to be released is not visible in the callback where the fix should be added.
	In these cases, \toolname's fix will also introduce a fresh variable in the same component where the leaked resource is \emph{acquired}, and make it point to the resource object.
	This ensures that a reference to the resource to be released is visible at the fix location.
	
	\fakepar{Fix injection.}
	A fix's resource release statement may be injected into the application at different locations.
	A simple, conservative choice would be the component's final callback function (\J{onDestroy} for activity components).
	Such a choice would be simple and functionally correct but very inefficient, since the app would hold the resource for much longer than it actually needs it.
	
	Instead, \toolname uses the information computed during leak analysis to find a suitable release location.
	As we discussed in \autoref{sec:callbacks}, the overall output of \toolname's leak analysis is
	an execution path that is leaking a certain resource.
	The path traverses a sequence $C_1; C_2; \ldots; C_n$ of callback functions determined by the component's callback graph,
	and is constructed by \toolname in a way that it ends with a call $C_n$ to the callback function where the resource may be released (according to the resource's API documentation).
	Therefore, \toolname adds the fix statement in callback $C_n$ just after the last usage of the resource in the callback (if there is any).
	
	In the running example of \autoref{fig:motivating}, \toolname inserts the call to \J{release} in callback \J{onPause}, which is as early as possible in the sequence of callbacks---as shown in the rightmost node of \autoref{fig:full-example}.
	
	\subsection{Validation}
	\label{sec:validation}
	
	Since leak analysis is sound (see \autoref{sec:limitations}),
	\toolname's fixes are correct by construction in the sense that they will remove the leak that is being repaired.
	However, since the resource release statement that fixes the leak is inserted in the first suitable callback (as described in \autoref{sec:fixgen}),
	it is possible that it interferes in unintended ways with other \emph{usages} of the resource.

	In order to determine whether its fixes may have introduced inconsistencies of this kind,
	\toolname performs a final \emph{validation} step, which
	runs a modified analysis that checks absence of new leaks as well as absence of use-after-release errors.
	This analysis reuses the techniques of \autoref{sec:intra-proc} and \autoref{sec:inter-proc} with the only twist that
	the pushdown automaton characterizing the property to be checked is now extended to also capture absence of use-after-release errors.
	
	\toolname's fixes release resources in the recommended callback function, but the app's developer may have ignored this recommendation and written code that still uses the resource in callbacks that occur later in the component's lifecycle.
	Such scenarios are the usual origin of failed validation:
	\toolname would fix the leak but it would also introduce a use-after-release error by releasing the resource too early.
	
	Continuing \autoref{fig:motivating}'s example of resource \J{MediaPlayer},
	suppose that the implementation does call the \J{release} operation
	but only in activity \J{onStop}.
	\toolname would still report a leak, since it does not find a \J{release} in
	\J{onPause}---which is where \J{MediaPlayer} should be released according to its
	API documentation.
	Accordingly, \toolname's fix for this leak would add \J{release} in \J{onPause}
	as described in \autoref{sec:example}.
	Validation of the leak would, however, fail because the programmer-written \J{release} in
	\J{onStop} introduces a double-release of the same resource, which
	validation detects.
	
	If validation fails, \toolname still outputs the invalid fix, which
	can be useful as a \emph{suggestion} to the developer---who
	remains responsible for modifying it in a way that doesn't conflict with the rest of the app's behavior. In particular,
	the experiments of \autoref{sec:answer-RQ1} confirm
	the intuition that validation fails when \toolname releases
	resources in the callback function recommended by the official resource documentation
	but the rest of the app's behavior conflicts with this recommendation.
	Therefore, the developer
	has the options of adjusting the fix or refactoring
	the app to follow the recommended guidelines.
	
	In the running example of \J{MediaPlayer}, 
	the programmer may either ignore \toolname's suggestion
	and keep releasing the resource in \J{onStop} (against the resource's recommendations),
	or accept it and remove the late call of \J{release} that becomes no longer necessary.
		
	Validation is an \emph{optional} step in \toolname.
	This is because it is not needed if we can assume that the app under repair follows Android's recommendation for when (in which callbacks) a resource should be used and released.
	As we will empirically demonstrate in \autoref{sec:experiments},
	validation is indeed usually not needed---but it remains available as an option in all cases where an additional level of assurance is required. 
	
	\subsection{Features and Limitations}
	\label{sec:limitations}
	
	Let us summarize \toolname's design features and the limitations of its current implementation.
	\autoref{sec:experiments} will empirically
	confirm these features and assess the practical impact of the limitations.
	
	\subsubsection{Soundness}
	A leak detection technique is \emph{sound}~\cite{ProgramAnalysisBook} 
	if, whenever it finds no leaks for a certain resource,
	it really means that no such leaks are possible in the app under analysis.
	
	\toolname's intra-procedural analysis is sound:
	it performs an exhaustive search of all possible paths,
	and thus it will report a leak if there is one.
	The inter-procedural analysis, however, has two possible sources of unsoundness.
	\begin{enumerate*}[label=(\alph*)]
		\item \label{unsound1} Since it performs a fixed-depth unrolling
		of paths in the callback graph (\autoref{sec:callbacks}),
		it may miss leaks that only occur along longer paths.
		
		\item \label{unsound2} Since it ranks procedures according to their call order (\autoref{sec:explicit-calls}),
		and such an order is not uniquely defined if the call graph has cycles,
		it may miss leaks that only occur in other procedure execution orders.
	\end{enumerate*}

	Both sources of unsoundness are unlikely to be a significant limitation in practice~\cite{soundiness}.
	A leak usually does not depend on the absolute number of times a resource is acquired or released,
	but only on whether acquires and releases are balanced.
	As long as we unroll each loop a sufficient number of times,
	unsoundness source \ref{unsound1} should
	not affect the analysis in practice.
	Furthermore, leak detection is monotonic with respect to the unrolling depth $D$:
	as $D$ increases, \toolname may find more leaks by exploring more paths, but
	a leak detected with some $D$ will also always be detected with a larger $D' > D$.
	
	As for the second source of unsoundness, 
	the Android development model, where the overall control flow is determined by implicit callbacks,
	makes it unlikely that user-defined procedures have circular dependencies.
	More precisely, \toolname's soundness is only affected by cycles in paths with acquire and release---not in plain application logic---and hence
	unsoundness source \ref{unsound2} is also unlikely to occur.
	
	The experiments of \autoref{sec:experiments} will confirm that \toolname is sound in practice
	by demonstrating that a wide range of Android applications trigger neither source of unsoundness.

	\subsubsection{Precision}
	A leak detection technique is \emph{precise}~\cite{ProgramAnalysisBook}
	if it never reports false alarms (also called false positives):
	whenever it detects a leak, that leak really occurs in some executions of the app under analysis.
	In the context of leak repair, many false alarms would generate many spurious fixes, which do not introduce bugs (since the analysis is sound)
	but are useless and possibly slow down the app.
	
	\toolname's analysis is, as is common for dataflow analyses,
	flow-sensitive but path-insensitive.
	This means that it over-ap\-prox\-i\-mates the paths that an app \emph{may} take without taking into account the feasibility of those paths.
	As a very simple example, consider a program that only consists of statement \J{if (false) res.a()},
	where \J{res} is a reference to a resource and \J{a} is an acquire operation.
	This program is leak free, since the lone acquire will never execute.
	However, \toolname would report a leak because it conservatively assumes that every branch is feasible.
	
	\emph{Aliasing} occurs when different references to the same resource may be available in the same app.
	Since \toolname does not perform alias analysis, this is another source of precision loss:
	a resource with two aliases \J{x} and \J{y} that is acquired using \J{x} and released using \J{y} will be considered leaking by \toolname,
	which thinks \J{x} and \J{y} are different resources.
	
	In practice, these two sources of imprecision are limitations to \toolname's applicability.
	When aliasing is not present,
	the experiments of \autoref{sec:experiments} indicate that the path-insensitive over-approximation built by \toolname is very precise in practice.
	Whether aliasing is present mostly depends on the kind of resource that is analyzed.
	As we discuss in \autoref{subjects},
	it is easy to identify resources
	whose usage is unlikely to introduce aliasing.
	\toolname is currently geared towards detecting and repairing leaks
	of such resources.
	
	A comprehensive empirical study of performance bugs in mobile apps~\cite{emseperformancebugs}
	found that about 36\%
	of the analyzed Android resource leaks
	were related to non-aliasing resources, such as \J{WiFi} and \J{Camera}, whose operations are
	energy intensive.
	The same study also reported that
	the leaks that affect these resources tend to have a much longer life than the memory-related leaks,
	as it takes developers longer, on average, to detect and fix the former over the latter.
	These data indicate that non-aliasing resource leaks are a significant fraction of the most common
	resource-related issues in Android apps, can have a negative impact on performance, and can be challenging
	to detect and fix. These observations motivate \toolname's focus on non-aliasing resources.
	
	\subsection{Implementation}
	\label{sec:implementation}
	
	We implemented \toolname in Python on top of \textsc{AndroGuard}~\cite{AndroGuard} and \textsc{Apktool}~\cite{Apktool}.
	
	\toolname uses \textsc{AndroGuard}---a framework to analyze Android apps---mainly
	to build the control-flow graphs of methods (which are the basis of our resource-flow graphs)
	and to process manifest files (extracting information about the components that make up an app).
	Using \textsc{AndroGuard} ensures that the control-flow graphs
	capture all behavior allowed by Java/Android (including, for example, exceptional control flow)
	in a simple form in terms of a few basic operations.
	
	\textsc{Apktool}---a tool for reverse engineering of Android apps---supports patch generation:
	\toolname uses it to decompile an app, 
	modify it with the missing release operations,
	and recompile the patched app back to executable format.
	\toolname's analysis and patching work on \emph{Smali} code---a human-readable format for the binary bytecode format \emph{DEX},
	which is obtained by decompiling from and compiling to the \emph{APK} format.
	
	\section{Experimental Evaluation}
	\label{sec:experiments}
	
	The overall goal of our experimental evaluation is
	to investigate whether \toolname is a practically viable approach for detecting and repairing resource leaks in Android applications.
	We consider the following research questions.
	\begin{description}[leftmargin=!,labelwidth=\widthof{\bfseries RQ55}]
		\item[\textbf{RQ1:}]  Does \toolname generate fixes that are \emph{correct} and ``safe''?
		\item[\textbf{RQ2:}]  How do \toolname's fixes compare to those written by \emph{developers}?
		\item[\textbf{RQ3:}]  Is \toolname \emph{scalable} to real-world Android apps?
		\item[\textbf{RQ4:}] How does \toolname \emph{compare} with other automated repair tools for
		Android resource leaks?
		\item[\textbf{RQ5:}]  How does \toolname's behavior depend on the \emph{unrolling depth} parameter, 
		which controls its analysis's level of detail? 
	\end{description}

	\begin{table*}
     \centering
     \begin{adjustwidth}{-4mm}{-4mm}
		\setlength{\tabcolsep}{3pt}
		\def\extraspace{\\[3pt]}
		\normalsize
		\begin{tabular}{rllcc}
			\toprule
			&  \multicolumn{2}{c}{\small\textsc{operations}} &
			\multicolumn{1}{c}{\small\textsc{released}} \\
			\cmidrule(rl){2-3}
			\multicolumn{1}{c}{\small\textsc{resource}} &
			\multicolumn{1}{c}{\small$a_k$} & \multicolumn{1}{c}{\small$r_k$} &
			\multicolumn{1}{c}{\small\textsc{callback}} &                                                 
			\multicolumn{1}{c}{\small\textsc{reentrant?}} \\
			\midrule
			\multirow{1}{*}{\J{AudioRecorder}} & {\J{new}} & {\J{release}} & \multirow{1}{*}{\J{onPause}, \J{onStop}} & \multirow{1}{*}{\textsc{n}} \extraspace
			\multirow{2}{*}{\J{BluetoothAdapter}} & {\J{enable}} & {\J{disable}} & \J{onStop} & \multirow{2}{*}{\textsc{n}} \\
			& {\J{startDiscovery}} & {\J{cancelDiscovery}} & \J{onPause}  \extraspace
			\multirow{3}{*}{\J{Camera}} & {\J{lock}} & {\J{unlock}} & \multirow{3}{*}{\J{onPause}} & \multirow{3}{*}{\textsc{n}} \\
			& {\J{open}} & {\J{release}} \\
			& {\J{startPreview}} & {\J{stopPreview}} \extraspace
			\multirow{1}{*}{\J{LocationListener}} & {\J{requestUpdates}} & {\J{removeUpdates}} & \multirow{1}{*}{\J{onPause}} & \multirow{1}{*}{\textsc{n}} \extraspace
			\multirow{2}{*}{\J{MediaPlayer}} & {\J{new}} & {\J{release}} & \multirow{2}{*}{\J{onPause}, \J{onStop}} & \multirow{2}{*}{\textsc{n}} \\
			& {\J{start}} & {\J{stop}} \extraspace
			\multirow{1}{*}{\J{Vibrator}} & {\J{vibrate}} & {\J{cancel}} & \multirow{1}{*}{\J{onDestroy}} & \multirow{1}{*}{\textsc{n}} \extraspace
			\multirow{1}{*}{\J{WakeLock}} & {\J{acquire}} & {\J{release}} & \multirow{1}{*}{\J{onPause}} & \multirow{1}{*}{\textsc{y}} \extraspace
			\multirow{1}{*}{\J{WifiLock}} & {\J{acquire}} & {\J{release}} & \multirow{1}{*}{\J{onPause}} & \multirow{1}{*}{\textsc{y}} \extraspace
			\multirow{1}{*}{\J{WifiManager}} & {\J{enable}} & {\J{disable}} & \multirow{1}{*}{\J{onDestroy}} & \multirow{1}{*}{\textsc{n}} \\
			\bottomrule 
		\end{tabular}
    \end{adjustwidth}
    \caption{Android resources analyzed with \toolname.
			For each \textsc{resource}, the table reports the acquire $a_k$ and release $r_k$ \textsc{operations} it supports (according to the resource's API documentation~\cite{AndroidAPIdocs}),
			the \textsc{callback} function \J{on...} where the resource should be \textsc{released} (according to the \emph{Android developer guides} \cite{AndoidActivityLifecycleGuidelines}),
			and whether the resource is \textsc{reentrant} (\textsc{y}es implies
			that absence of leaks is a context-free property and \textsc{n}o implies that it is a regular property).
		}
		\label{tab:resources}
	\end{table*}
	
	\begin{table}
		\centering
		\setlength{\tabcolsep}{7pt}
		\def\extraspace{\\[3pt]}
		\normalsize
		\begin{tabular}{lrlr}
			\toprule
			\multicolumn{1}{c}{\textsc{app}} & \multicolumn{1}{c}{\textsc{kloc}} & \multicolumn{1}{c}{\textsc{resource}} & \multicolumn{1}{c}{\textsc{leaks}} \\
			\midrule
			{APG}  &  {42.0}  &  {\footnotesize\J{MediaPlayer}}  &  1 \extraspace
			{BarcodeScanner}  &  {10.6}  &  {\footnotesize\J{Camera}}  &  1  \extraspace
			{CallMeter}  &  {13.5}  &  {\footnotesize\J{WakeLock}}  &  3 \extraspace
			\multirow{2}{*}{ChatSecure}  &  \multirow{2}{*}{37.2}  &  {\footnotesize\J{BluetoothAdapter}}  &  0 \\
			&     &  {\footnotesize\J{Vibrator}}  &  1 \extraspace
			{ConnectBot}  &  {17.6}  &  {\footnotesize\J{WakeLock}}  &  0 \extraspace
			{CSipSimple}  &  {49.0}  &  {\footnotesize\J{WakeLock}}  &  2 \extraspace
			\multirow{2}{*}{IRCCloud}  &  \multirow{2}{*}{35.3}  &  {\footnotesize\J{MediaPlayer}}  &  0 \\
			&     &  {\footnotesize\J{WifiLock}}  &  1 \extraspace
			{K-9 Mail}  &  {78.5}  &  {\footnotesize\J{WakeLock}}  &  2 \extraspace
			{OpenGPSTracker}  &  {12.3}  &  {\footnotesize\J{LocationListener}}  &  1 \extraspace
			{OsmDroid}  &  {18.4}  &  {\footnotesize\J{LocationListener}}  &  2 \extraspace
			{ownCloud}  &  {31.6}  &  {\footnotesize\J{WifiLock}}  &  2 \extraspace
			{QuranForAndroid}  &  {21.7}  &  {\footnotesize\J{MediaPlayer}}  &  1 \extraspace
			{SipDroid}  &  {24.5}  &  {\footnotesize\J{Camera}}  &  4 \extraspace
			{SureSpot}  &  {41.0}  &  {\footnotesize\J{MediaPLayer}}  &  2 \extraspace
			{Ushahidi}  &  {35.7}  &  {\footnotesize\J{LocationListener}}  &  1 \extraspace
			{VLC}  &  {18.1}  &  {\footnotesize\J{WakeLock}}  &  2 \extraspace
			{Xabber}  &  {38.2}  &  {\footnotesize\J{AudioRecorder}}  &  2 \extraspace
			\midrule
			\multicolumn{1}{c}{\textsc{average}} & 30.9 \\
			\multicolumn{1}{c}{\textsc{total}}  &  525.2  &     &  26 \\
			\bottomrule
		\end{tabular}
		\caption{\leaks apps analyzed with \toolname.
			For each \textsc{app}, the table reports its size \textsc{kloc} in thousands of lines of code.
			For each \textsc{resource} used by the app,
			the table then reports the number of \textsc{leaks} of that resource and app included in \leaks.
			The two bottom rows report the \textsc{average} (mean) and \textsc{total}
			for all apps.
		}
		\label{tab:benchmark}
	\end{table}

	\subsection{Experimental Setup}
	\label{sec:setup}
	
	This section describes how we selected the apps used in the experimental
	evaluation of \toolname, how we ran the experiments, and how we collected
	and assessed the experiments' results to answer the research questions.
	
	\subsubsection{Subjects: RQ1, RQ2, RQ3, RQ5}
	\label{subjects}
	Our experiments to assess correctness and scalability target apps in \leaks~\cite{droidleaks}---a curated collection of resource leak bugs in real-world Android applications.
	\leaks collects a total of 292 leaks from 32 widely used open-source Android apps.
	For each leak, \leaks includes both the buggy (leaking) version of an app and a leak-free version obtained by manually fixing the leak.

	Leaks in \leaks affect 22 resources. 
	The majority of them (13) are Android-specific resources (such as \J{Camera} or \J{WifiLock}),
	while the others are standard Java APIs (such as \J{InputStream} or \J{BufferReader}).
	\toolname's analysis is based on the Android programming model,
	and every Android-specific resource expresses its usage policy in terms of the callback functions where a resource can be acquired or released---an
	information that is not available for standard Java API's resources.
	Therefore, our evaluation only targets leaks affecting Android-specific resources. 
	As we discussed in \autoref{sec:limitations},
	\toolname is oblivious of possible aliases between references to the same resource object.
	If such aliasing happens within the same app's implementation,
	it may significantly decrease \toolname's precision.
	We found that each Android resource can naturally be classified into \emph{aliasing} and \emph{non-aliasing}
	according to whether typical usage of that resource in an app may introduce multiple references that alias one another.\footnote{%
		Note that the classifications of resources into aliasing/non-aliasing and reentrant/non-reentrant are orthogonal:
		\toolname fully supports reentrant resources, but achieves a high precision only when analyzing non-aliasing resources.
	}
	Usually, a non-aliasing resource is one that is accessed in strict mutual exclusion,
	and hence such that obtaining a handle is a relatively expensive operation;
	\J{Camera}, \J{MediaPlayer}, and \J{AudioRecorder} are examples of non-aliasing resources.
	In contrast, aliasing resources tend to 
	support a high degree of concurrent access,
	and hence it is common to instantiate fresh handles for each usage;
	a database \J{Cursor} is a typical example of such resources,
	as creating a new cursor is inexpensive, and database systems support fine-grained concurrent access.
	Out of all 13 Android resources involved in leaks in \leaks,
	9 are non-aliasing;
	our experiments ran \toolname on all apps in \leaks that use these resources.\footnote{
		We also tried \toolname on the 4 ($= 13 - 9$) aliasing resources in \leaks;
		\autoref{sec:aliasing} discusses the outcome of these secondary experiments.
	}
	
	\autoref{tab:resources} summarizes the characteristics of the 9 resources we selected for our experiments according to the above criteria.
	Then, \autoref{tab:benchmark} lists all apps in \leaks that use some of these resources,
	their size, and how many leaks of each resource \leaks includes with fixes.
	Thus, the first part of our experiments will
	target 16 Android apps with a total size of about half a million lines of code;
	\leaks collects 26 leaks in these apps affecting the 9 non-aliasing resources we consider.
	
	\subsubsection{Subjects: RQ4, RQ5}
	\label{subjects-comparison}
	
	At the time of writing, RelFix~\cite{Relfix} is the only fully automated tool
	for the detection and repair of Android resource leaks,
	which can be quantitatively compared to \toolname.
	Since RelFix uses Relda2~\cite{Relda2TSE16} to \emph{detect} leaks, and Relda2's
	experimental evaluation is broader than RelFix's, we also compare \toolname's leak
	detection capabilities to Relda2's.
	
	As we discuss in \autoref{sec:related-leaks}, other tools exist
	that detect other kinds of leaks; since they are not directly applicable
	to the same kinds of resources that \toolname analyzes, we only compare them
	to \toolname in a qualitative way in \autoref{sec:relatedwork}.
	
	The experimental evaluations of RelFix and Relda2,
	as reported in their publications~\cite{Relfix,Relda2TSE16}, targeted 27
	Android apps that are not part of \leaks.
	Unfortunately, neither detailed experimental results (such as the app versions
	that were targeted or the actual produced fixes) nor the RelFix and Relda2
	tools are publicly available.\footnote{%
		The authors of \cite{Relda2TSE16,Relfix} could not follow up on our requests to share
		the tools or details of their experimental evaluation.
	}
	Therefore, a detailed, direct experimental comparison is not possible.
	However, we could still run \toolname on the same apps analyzed with RelFix and Relda2, and
	compare our results to those reported by \cite{Relfix,Relda2TSE16}
	in terms of number of fixes and precision.
	
	Out of the 27 apps used in RelFix and Relda2's experimental evaluations,
	22 use some of the 9 non-aliasing resources
	that \toolname targets (see \autoref{tab:resources}).
	More precisely, Relda2's evaluation in~\cite{Relda2TSE16} \emph{only} targets resources that are
	non-aliasing, thus we consider all of their experiments in our comparison.
	RelFix's evaluation in~\cite{Relfix} targets 4 non-aliasing and 4 aliasing resources;
	we only consider the former for our comparison with \toolname.
	In order to include apps
	as close as possible to those actually analyzed by \cite{Relfix,Relda2TSE16}, we downloaded
	the \texttt{apk} release of each app that was closest in time to the publication time of \cite{Relfix,Relda2TSE16}.
	This excluded 2 apps whose older releases we could not retrieve.
	In all, this process identified 20 apps: 16 used in the evaluation of Relda2 (listed in \autoref{tab:relda2}) and 4 used in the evaluation of RelFix (listed in \autoref{tab:relfix}),
	for a total of \numprint{260000} lines of code and using 7 of the 9 non-aliasing resources.
	
	Relda2 supports both a flow-insensitive and a flow-sensitive detection algorithm.
	According to \cite{Relda2TSE16}, the flow-insensitive approach is faster but
	much less precise.
	Since \toolname's analysis is also flow-sensitive, we only compare it to
	Relda2's flow-sensitive analysis (option~$\mathit{op}_{2}$ in \cite{Relda2TSE16}),
	which is also the one used by RelFix.
	
	\subsubsection{Experimental Protocol}
	\label{sec:ex-protocol}
	
	In our experiments,
	each run of \toolname targets one app and repairs leaks of a specific resource.\footnote{
		\toolname can analyze leaks for multiple resources in the same run, 
		but we do not use this features in the experiments in order to have a fine-grained breakdown of \toolname's performance.}
	The run's output is a number of leaks and, for each of them, a fix.
	
	After each run, we manually inspected the fixes produced by \toolname,
	confirmed that they are syntactically limited to a small number of release operations,
	and checked that the app with the fixes still runs normally.
	Unfortunately, the apps do not include tests that we could have used as additional evidence
	that the fixes did not introduce any regression.
	However, \toolname's soundness guarantees that the fixes are correct by construction;
	and its validation phase further ascertains that the fixes
	do not introduce use-after-release errors.
	
	In all experiments with \leaks,
	we also tried to match, as much as possible, the leaks detected and repaired by \toolname
	to those reported in \leaks.
	This was not always possible: some apps' are only available in obfuscated form,
	which limits what one can conclusively determine by inspecting the bytecode.
	In addition, \leaks's collection is not meant to be exhaustive: therefore,
	it is to be expected that \toolname finds leaks that are not included in \leaks.
	In the experiments with the apps analyzed by RelFix and Relda2, we did not have any ``ground truth'' to compare them to, but we still performed manual checks and testing.
	
	More precisely, we followed these steps to manually inspect and validate
	all leak detected by \toolname:
	\begin{enumerate*}
		\item We consider the leaking path on the resource-flow graph reported by \toolname
		and determine whether it is feasible. If this is not the case (for example, two elements of the path condition are contradictory), then we classify the leak as a false positive.
		\item If the leaking path is feasible, we consider the path's matching sequence of callbacks, and use that
		as a guide to write a test that tries to cover the path on the real program---and thus confirms that the leak is a
		true positive.
		For short paths, the path condition usually contains enough information to come up with a test
		after some trial-and-error.
		\item 
		In more complex cases (especially paths involving UI interactions),
		we use Android Studio's \emph{Monkey Testing} framework~\cite{MonkeyTester}
		to generate several random input sequences that thoroughly exercise the app;
		then, we monitor the app running on those inputs, and select any execution paths that
		matched the leaking path on the resource-flow graph.
		\item Finally, we re-compile the app after injecting the fix produced by \toolname, and
		run the patched app on the tests generated as described above, as well as on a few other random inputs
		and also trying the app interactively,
		checking that the leak is no longer triggered and there are no other changes in behavior
		(in particular, no negative impact on performance).
	\end{enumerate*}
	In all cases, we were conservative in assessing which leaks and fixes are correct,
	marking as ``confirmed'' only cases where
	the collected evidence that a leak may occur and its fix safely repairs it
	is conclusive.
	
	The main parameter regulating \toolname's behavior is the unrolling depth $D$.
	We ran experiments with $D$ ranging from $1$ to $6$, to demonstrate empirically
	that the default value $D=3$
	is necessary and sufficient to achieve soundness (i.e., no leaks are missed). 
	
	\fakepar{Hardware/software setup.}
	All the experiments ran on a MacBook Pro equipped with
	a 6-core Intel Core i9 processor and 16~GB of RAM,
	running macOS~10.15.3,
	Android 8.0.2 with API~level~26, Python~3.6, \textsc{AndroGuard} 3.3.5, \textsc{Apktool} 2.4.0.
	
	\begin{table*}[!tb]
     \centering
     \small
		\setlength{\tabcolsep}{1.5pt}
		\def\extraspace{\\[4pt]}
      \begin{adjustwidth}{-25mm}{-10mm}
		\begin{tabular}{lrrrlrrrrrrrr}
			\toprule
			&\multicolumn{2}{c}{$\textsc{rfg}$} & \multicolumn{1}{c}{\textsc{cc}} & & \multicolumn{5}{c}{\textsc{time} (s)} & \multicolumn{1}{c}{} \\
			\cmidrule(lr){2-3} \cmidrule(lr){6-10}
			\multicolumn{1}{c}{\textsc{app}} &  \multicolumn{1}{c}{$|V|$} & \multicolumn{1}{c}{$|E|$} & \multicolumn{1}{c}{$M/M'$}& \multicolumn{1}{c}{\textsc{resource}} & \multicolumn{1}{c}{\textsc{abstraction}} & \multicolumn{1}{c}{\textsc{analysis}} & \multicolumn{1}{c}{\textsc{fixing}} & \multicolumn{1}{c}{\textsc{validation}} & \multicolumn{1}{c}{\textsc{total}} & \multicolumn{1}{c}{\textsc{fixed}} & \multicolumn{1}{c}{\checkQM} & \multicolumn{1}{c}{\textsc{invalid}} \\
			\midrule
			{APG}  & {\numprint{4968}} & {\numprint{7442}} & {0.47} & {\footnotesize\J{MediaPlayer}} & 32.7 & 273.4 & 0.2 & 55.5 & 361.8 & 1 & 0 & 0
			\extraspace
			
			{BarcodeScanner} & {\numprint{1189}} & {\numprint{2462}} & {0.35} & {\footnotesize\J{Camera}} & 6.8 & 69.2 & 0.3 & 21.8 & 98.1 & 3 & 0 & 0
			\extraspace
			
			{CallMeter} & {\numprint{1840}} & {\numprint{3216}} & {0.35} & {\footnotesize\J{WakeLock}} &11.2 & 100.7 & 0.6 & 35.1 & 147.6 & 4 & 0 & 0
			\extraspace
			
			\multirow{2}{*}{ChatSecure} & \multirow{2}{*}{\numprint{5430}} & \multirow{2}{*}{\numprint{8686}} & \multirow{2}{*}{0.48} & {\footnotesize\J{BluetoothAdapter}} & \multirow{2}{*}{23.0}  & 316.6 & 0.1 & 114.6 & 454.3 & 2 & 0 & 0
			\\
			
			&   &   &   & {\footnotesize\J{Vibrator}} &   & 273.7 & 0.5 & 93.3 & 390.2 & 2 & 0 & 0
			\extraspace
			
			{ConnectBot} & {\numprint{1956}} & {\numprint{3814}} & {0.23} & {\footnotesize\J{WakeLock}} & 12.1 & 107.0 & 0.2 & 32.7 & 152.0 & 2 & 0 & 0
			\extraspace
			
			{CSipSimple} & {\numprint{5712}} & {\numprint{9154}} & {0.42} & {\footnotesize\J{WakeLock}} & 38.0 & 433.8 & 0.1 & 111.4 & 583.3 & 4 & 2 & 0
			\extraspace
			
			\multirow{2}{*}{IRCCloud} & \multirow{2}{*}{\numprint{4782}} & \multirow{2}{*}{\numprint{9755}} & \multirow{2}{*}{0.43} & {\footnotesize\J{MediaPlayer}} & \multirow{2}{*}{25.1} & 239.8 & 0.5 & 80.5 & 345.9 & 3 & 0 & 0
			\\
			
			&   &   &   & {\footnotesize\J{WifiLock}} &   & 295.3 & 0.3 & 58.0 & 380.8& 2 & 0 & 0
			\extraspace
			
			{K-9 Mail} & {\numprint{8831}} & {\numprint{16390}} & {0.29} & {\footnotesize\J{WakeLock}} & 64.1 & 475.7 & 0.4 & 165.8 & 706.0 & 2 & 0 & 0
			\extraspace
			
			{OpenGPSTracker} & {\numprint{1418}} & {\numprint{2791}} & {0.29} & {\footnotesize\J{LocationListener}} & 7.1 & 107.4 & 0.4 & 33.7 & 148.6 & 2 & 0 & 0
			\extraspace
			
			{OsmDroid} & {\numprint{2222}} & {\numprint{3545}} & {0.36} & {\footnotesize\J{LocationListener}} & 12.9 & 161.9 & 0.4 & 32.9 & 208.1 & 4 & 2 & 0
			\extraspace
			
			{ownCloud} & {\numprint{4444}} & {\numprint{8980}} & {0.6} & {\footnotesize\J{WifiLock}} & 20.7 & 238.6 & 0.5 & 47.8 & 307.6 & 4 & 2 & 0
			\extraspace
			
			{QuranForAndroid} & {\numprint{2898}} & {\numprint{4545}} & {0.43} & {\footnotesize\J{MediaPlayer}} & 14.9 & 177.5 & 0.2 & 63.8 & 256.4 & 2 & 0 & 0
			\extraspace
			
			{SipDroid} & {\numprint{3178}} & {\numprint{4583}} & {0.38} & {\footnotesize\J{Camera}} & 14.1 & 176.6 & 0.5 & 39.7 & 230.9 & 4 & 0 & 0
			\extraspace
			
			{SureSpot} & {\numprint{3575}} & {\numprint{7240}} & {0.37} & {\footnotesize\J{MediaPLayer}} & 33.0 & 246.4 & 0.4 & 54.1 & 333.9 & 3 & 0 & 3
			\extraspace
			
			{Ushahidi} & {\numprint{5073}} & {\numprint{10417}} & {0.43} & {\footnotesize\J{LocationListener}} & 24.4 & 201.9 & 0.3 & 58.5 & 285.1 & 2 & 0 & 2
			\extraspace
			
			{VLC} & {\numprint{2689}} & {\numprint{4199}} & {0.55} & {\footnotesize\J{WakeLock}} & 14.4 & 119.5 & 0.5 & 34.4 & 168.8 & 2 & 0 & 0
			\extraspace
			
			{Xabber} & {\numprint{4194}} & {\numprint{8478}} & {0.31} & {\footnotesize\J{AudioRecorder}} &25.4 & 256.9 & 0.3 & 76.6 & 359.2& 2 & 0 & 0
			\extraspace
			
			\midrule
			
			\multicolumn{1}{c}{\textsc{average}} & \numprint{3788} & \numprint{6805} & 0.4 &   & 23.4 & 231.2 & 0.3 & 65.2 & 320.2 &   &  
			\extraspace
			\multicolumn{1}{c}{\textsc{total}} & \numprint{64399} & \numprint{115697} & 6.74 &   & 444.6 & 4392.8 & 5.9 & 1238.8 & 6084.8 & 50 & 6 & 5
			\\
			\bottomrule
		\end{tabular}
    \end{adjustwidth}
    \caption{Results of running \toolname on apps in \leaks.
			For every \textsc{app}, the table reports 
			the number of nodes $|V|$ and edges $|E|$ of its resource-flow graph \textsc{rfg},
			and the ratio $M/M'$ between the \textsc{rfg}'s cyclomatic complexity $M$ and the cyclomatic complexity  $M'$
			of the whole app's control-flow graph.
			For every \textsc{resource} used by the app,
			the table then reports 
			\toolname's running time
			to perform each of the steps of \autoref{fig:workflow} (\textsc{abstraction}, \textsc{analysis}, \textsc{fixing}, and \textsc{validation});
			as well as the \textsc{total} running time;
			since the abstraction is built once per app, time \textsc{abstraction}
			is the same for all resources used by an app.
			Finally, the table reports the number of leaks of each resource detected and \textsc{fixed} by \toolname;
			how many of these fixed leaks we could not conclusively classify as real leaks (\checkQM);
			and the number of the fixes that \toolname classified as \textsc{invalid} (that is, they failed validation).
			The two bottom rows report the \textsc{average} (mean, per app or per app-resource) and \textsc{total} in all experiments.
		}
		\label{tab:results}
	\end{table*}
		
	\subsection{Experimental Results}
	\label{sec:results}

	This section summarizes the results of our experiments with \toolname, and
	discusses how the results answer the research questions.
	
	\subsubsection{RQ1: Correctness}
	\label{sec:answer-RQ1}
	Column \textsc{fixed} in \autoref{tab:results} reports the number of \leaks leaks
	that \toolname detected and fixed with a correct fix (i.e., a fix that prevents leaking);
	column \textsc{invalid} reports how many of these fixes failed validation (were ``unsafe'').
	\toolname was very effective at detecting leaks in non-aliasing resources.
	In particular, it detected and fixed \emph{all} 26 leaks reported by \leaks and included in our experiments (see \autoref{tab:benchmark}), building a correct fix for each of them.
	In addition, it detected and fixed another 24 leaks in the same apps.
	
	\nicepar{Precision.}
	Empirically evaluating \emph{precision} is tricky because we lack a complete baseline.
	By design, \leaks is not an \emph{exhaustive} collection of leaks.
	Therefore, when \toolname reports and fixes a leak it could be:
	\begin{enumerate*}[label=(\alph*)]
		\item \label{l:inDL} a real leak included in \leaks;
		\item \label{l:tp} a real leak \emph{not} included in \leaks;
		\item \label{l:fp} \label{leak:spurious} a spurious leak.
	\end{enumerate*}
	By inspecting the leak reports and the apps (as discussed in \autoref{sec:ex-protocol})
	we managed to 
	confirm that 44 leaks (88\%) reported by \toolname are in categories \ref{l:inDL}
	(26 leaks or 52\%, matching all leaks included in \leaks)
	or \ref{l:tp} (18 leaks or 36\%) above---and
	thus are real leaks (true positives).
	Unfortunately, the remaining 6 leaks (12\%) reported by \toolname
	were found in apps whose bytecode is only available in \emph{obfuscated} form,
	which means we cannot be certain they are not spurious;
	these unconfirmed cases are counted in column ``\checkQM'' in \autoref{tab:results}.
	Even in the worst case in which all of these are spurious,
	\toolname's precision would remain high (88\%).
	The actual precision is likely higher: 
	in all cases where we could analyze the code, we found a real leak;
	unconfirmed cases probably just require more evidence
	such as access to unobfuscated bytecode.
	Anyway, note that any spurious fixes would still be safe to apply---albeit unnecessary---because they do not introduce bugs: since all release operations added by PlumbDroid are conditional (\autoref{sec:fixgen}),
	a fix ``repairing'' a spuriously detected leak simply introduces release operations that is neve executed in actual program executions.
		
	\nicepar{Correctness and safety.}
	All fixes built by \toolname are correct in the sense that they release resources so as to avoid a leak;
	manual inspection, carried out as described in \autoref{sec:ex-protocol}, 
	confirmed this in all cases---with some remaining uncertainty only for obfuscated apps.
	
	\toolname's validation step assesses ``safety'': whether a fix does not introduce a use-after-release error.
	All but 5 fixes built by \toolname for non-aliasing resources are safe.
	The 5 unsafe fixes are:
	\begin{enumerate}[label=(\roman*)]
		\item Three identical fixes (releasing the same resource in the same location)
		repairing three distinct leaks of resource \J{MediaPlayer} in app SureSpot.
		
		According to the Android reference manual~\cite{AndroidRef}, this resource can be released either in the \J{onPause} or in the \J{onStop} callback.
		\toolname releases resources as early as possible by default, and hence it built a fix releasing the \J{MediaPlayer} in \J{onPause}.
		The developers of SureSpot, however, assumed that the resource is only released later (in \J{onStop}), and hence \toolname's fix
		introduced a use-after-release error that failed validation.
		
		To deal with such situations---resources that may be released in different call\-backs---we
		then could introduce a configuration option to decide whether to release resources \emph{early} or \emph{late}.
		An app developer could therefore configure our analyzer in a way that suits their design decisions.
		In particular, configuring \toolname with option \emph{late} in these cases would generate fixes that pass validation.
		
		\item Two identical fixes (releasing the same resource in the same location) repairing two distinct leaks of resource \J{LocationListener} in app Ushahidi.
		
		The fix generated by \toolname failed validation because 
		the app's developers assumed that the resource is only released in callback \J{onDestroy}.
		This assumption conflicts with Android's recommendations to release the resources in earlier callbacks.
		
		In this case, the best course of action would be amending the app's usage policy of the resource so as to comply with Android's guidelines.
		\toolname's fix would pass validation after this modification.
	\end{enumerate}
	
	Note that \toolname does not output fixes that do not pass the validation step;
	therefore, there is no risk that such unsafe fixes are accidentally deployed.
	
	\begin{result}
		\toolname detected and fixed 50 leaks in \leaks producing correct-by-construction fixes.
		\toolname's detection is very precise on the resources it supports.
	\end{result}
	
	\begin{figure}[!tb]
		\begin{lstlisting}
public class NetworkConnection {

    public void disconnect() {
        if (client != null) {
            state = STATE_DISCONNECTING;
            client.disconnect();
        } else {
            state = STATE_DISCONNECTED;
        }
        if (idleTimer != null) {
            idleTimer.cancel();
            idleTimer = null;
        }
        (*\label{l:case1-patch}\textbf{\textcolor{vorange}{
        if (wifiLock.isHeld()) wifiLock.release();
         }} *)
    }

}
		\end{lstlisting} 
		\caption{An excerpt of class \J{NetworkConnection} in Android app IRCCloud, showing the \J{disconnect()} method
			with a resource leak (code in black, executed in callback \J{onPause()}),
			and how the leak was patched by the app developers in commit \href{https://github.com/irccloud/android/commit/113555e9ae7a2f4b8ec83b4d2e17729266c9c8d4}{113555e9ae}
			(code in \textbf{\textcolor{vorange}{orange}}).
			\toolname generates a patch for this leak that exactly matches the developer-written one shown here.}
		\label{fig:case1}
	\end{figure}

	\begin{figure*}[!tb]
     \centering
     \begin{adjustwidth}{-12mm}{-12mm}
		\begin{subfigure}[t]{0.62\textwidth}
			\begin{lstlisting}[basicstyle=\ttfamily\footnotesize]
public class NetworkConnection  {
			
  public void removeHandler(Handler handler) {
    handlers.remove(handler);
    if (handlers.isEmpty()) {
      if (shutdownTimer == null) {
        // ...
        disconnect();
      }
      if (idleTimer != null 
          && state != STATE_CONNECTED) {
        idleTimer.cancel();
        idleTimer = null;
        failCount = 0;
        (*\label{l:case2-developer}\textbf{\textcolor{vorange}{
        if (wifiLock.isHeld()) wifiLock.release();
        }} *)
        reconnect_timestamp = 0;
        state = STATE_DISCONNECTED;
      }
    }
  }

}
			\end{lstlisting} 
			\caption{The app developers fixed the leak in commit \href{https://github.com/irccloud/android/commit/35e0a587e3e9ed376b36355dfbccdeed049aae85}{35e0a587e3}
				\newline by adding a conditional release at line~\ref{l:case2-developer} (code in \textbf{\textcolor{vorange}{orange}}).}
			\label{fig:case2-developer}
		\end{subfigure}
		\hfill
		\begin{subfigure}[t]{0.62\textwidth}
			\begin{lstlisting}[basicstyle=\ttfamily\footnotesize]
public class NetworkConnection  {
			
  public void removeHandler(Handler handler) {
    handlers.remove(handler);
    if (handlers.isEmpty()) {
      if (shutdownTimer == null) {
        // ...
        disconnect();
      }
      if (idleTimer != null 
          && state != STATE_CONNECTED) {
        idleTimer.cancel();
        idleTimer = null;
        failCount = 0;

        reconnect_timestamp = 0;
        state = STATE_DISCONNECTED;
        (*\label{l:case2-plumbdroid}\textbf{\textcolor{vgreen}{
			if (wifiLock.isHeld()) wifiLock.release();
			}} *)
      }
    }
  }

}
			\end{lstlisting} 
			\caption{\toolname fixed the leak by adding a conditional release at line~\ref{l:case2-plumbdroid} (code in \textbf{\textcolor{vgreen}{green}}). \toolname only detects this leak with unrolling depth $D \geq 2$,
				as it affects re-entrant resource \J{WifiLock}.}
			\label{fig:case2-plumbdroid}
       \end{subfigure}
     \end{adjustwidth}
     \caption{An excerpt of class \J{NetworkConnection} in Android app IRCCloud, showing the  \J{removeHandler()} method
			with a resource leak (code in black, executed in callback \J{onPause()}), and how it was fixed by the app developers (\autoref{fig:case2-developer}) and by \toolname (\autoref{fig:case2-plumbdroid}).}
		\label{fig:case2}
	\end{figure*}

	\subsubsection{RQ2: Comparison with Developer Fixes}
	\label{sec:answer-RQ2}
	
	The manual inspection and validation protocol (\autoref{sec:ex-protocol}) that we followed to
	confirm the correctness of all leaks and fixes reported by \toolname
	already strongly suggests that the results of \toolname's analysis are usually of high quality.
	To better understand whether \toolname's fixes are comparable to those written by developers,
	we further scrutinized the fixes produced by \toolname for leaks in apps IRCCloud and Ushahidi.
	We selected these two apps as they are among the largest in the \droidl collection,
	they are open source, and their public code repositories are well organized and feature
	a long-running development history. Furthermore, as shown in \autoref{tab:results},
	\toolname detected and fixed 5 leaks in IRCCloud; and detected 2 leaks in Ushahidi, for which
	it could only produce fixes that fail the validation step. Thus, there is a mix of cases
	where \toolname is completely successful and cases where \toolname's validation fail---which
	mitigates the risk of biasing the analysis in \toolname's favor.
	
	Based on these results, we went through the bug reports and commit history of the two apps,
	looking for developer-written fixes of the 7 leaks that \toolname detected and fixed.
	Indeed, we found that developers eventually found and fixed all these leaks,\footnote{See IRCCloud's commits: \href{https://github.com/irccloud/android/commit/113555e9ae7a2f4b8ec83b4d2e17729266c9c8d4}{113555e9ae},  \href{https://github.com/irccloud/android/commit/35e0a587e3e9ed376b36355dfbccdeed049aae85}{35e0a587e3}, 
		\href{https://github.com/irccloud/android/commit/0cd91bc5ca350671bcf6ae84d634d097a3602d8c}{0cd91bc5ca}, 
		\href{https://github.com/irccloud/android/commit/d7a441e3a675cac30cffdfdfa94e5a6dd486b169}{d7a441e3a6}, 
		\href{https://github.com/irccloud/android/commit/113555e9ae7a2f4b8ec83b4d2e17729266c9c8d4}{113555e9ae};
		and Ushahidi's commits:
		\href{https://github.com/ushahidi/Ushahidi_Android/commit/9d0aa75b84d74566727b91f5d7dcb85caff34d33}{9d0aa75b84}, 
		\href{https://github.com/ushahidi/Ushahidi_Android/commit/337b48f5f2725f3e84796fab12947ffbec3c0357}{337b48f5f2}.}
	which confirms that they are considered serious enough faults.
	
	In 2 out of the 5 leaks in IRCCloud,
	the fix produced by \toolname is identical to the developer-written one,
	as they both release the same resource in the same location.
	\autoref{fig:case1} shows one of these leaks and its fix.\footnote{In these examples, we express the fix as Java source code, even though \toolname works at the level of the human-readable bytecode Smali.}
	In the other 3 leaks in IRCCloud,
	the fix produced by \toolname released the same resource as the developer-written one
	but in a different location of the same basic block.
	\autoref{fig:case2} shows one of these leaks and its fix by developer (\autoref{fig:case2-developer})
	and by \toolname (\autoref{fig:case2-plumbdroid}).
	In all these cases, the difference of release location is immaterial, as the statements that are
	between the release point in the developer-written fix and the release point in \toolname's fix
	are simple assignments do not affect any shared resources and execute quickly.
	
	In both leaks in Ushahidi,
	the difference between the (invalid) fixes produced by \toolname and
	the developer-written ones is also the location of the release.
	Sticking to the Android developer guidelines~\cite{AndoidActivityLifecycleGuidelines},
	\toolname only releases the resource \linebreak\J{LocationListener} in the callback \J{onPause()} for one of these leaks;
	doing so fails the validation phase,
	since the Ushaidi app still uses location resources when it is running in the background.
	Therefore, the developers fixed the leak by releasing the resource
	in the callback \J{onDestroy()}, that is only when the app is shut down.
	The other leak has a similar discrepancy between the Android guidelines followed by \toolname
	and how the Ushahidi app is written.
	
	In all, \toolname's fixes are often very similar and functionally equivalent
	to those written by programmers for the same leaks.
	\toolname's validation phase is useful to detect when an app deviates from 
	Android's guidelines on resource management;
	in these cases, the user of \toolname can decide whether to refactor the app
	to follow the guidelines, or modify \toolname so that it generates a fix at a different location.
	
	The similarity between \toolname and developer-written fixes, as well their usual syntactic simplicity,
	is also evidence that these fixes are unlikely to negatively alter the running-time performance
	of an app (and hence the user experience). This is also consistent with our manual analysis (\autoref{sec:ex-protocol}),
	which never found an app's responsiveness to worsen after applying a resource-leak fix.
	
	\begin{result}
		According to the analysis of a sample, we found that the fixes produced by \toolname 
		often are functionally equivalent to those written by the app developers.
	\end{result}

	\subsubsection{RQ3: Performance}
	\label{sec:answer-RQ3}
	
	Columns \textsc{time} in \autoref{tab:results} report the running time of \toolname in each step.
	As we can expect from a tool based on static analysis,
	\toolname is generally fast and scalable on apps of significant size.
	Its average running time is around 5 minutes \emph{per app-resource} and around 2 minutes \emph{per repaired leak} ($121\,\mathrm{s} \simeq 6084.8 / 50$).
	
	The \textsc{analysis} step dominates the running time, since it performs an exhaustive search.
	In contrast, the \textsc{abstraction} step is fairly fast (as it amounts to simplifying control-flow graphs);
	and the \textsc{fixing} step takes negligible time (as it directly builds on the results of the analysis step).
	
	\toolname's abstractions are key to its performance, as we can see from \autoref{tab:results}'s data about the size of the resource-flow graphs.
	Even for apps of significant size, the resource-flow graph remains manageable;
	more important, its \emph{cyclomatic complexity} $M$---a measure of the number of paths in a graph~\cite{McCabe76}---is usually much lower than the cyclomatic complexity  $M'$ of the full control-flow graph,
	which makes the exhaustive analysis of a resource-flow graph scalable.
	
	\begin{result}
		\toolname is scalable: it takes about 2 minutes \\on average to detect and fix a resource leak.
	\end{result}
	
	\begin{table*}
		\centering
		\def\extraspace{\\[3pt]}
		\normalsize
      \begin{adjustwidth}{-4mm}{-4mm}
		\begin{tabular}{lrlrrrrr}
			\toprule
			& & & \multicolumn{3}{c}{\toolname} & \multicolumn{2}{c}{Relda2} \\
			\cmidrule(lr){4-6} \cmidrule(lr){7-8}
			\multicolumn{1}{c}{\textsc{app}} & \multicolumn{1}{c}{\textsc{kloc}} & \multicolumn{1}{c}{\textsc{resource}} & \multicolumn{1}{c}{\textsc{time} (s)} &
			\multicolumn{1}{c}{\textsc{leaks}} & \multicolumn{1}{c}{\checkOK} &
			\multicolumn{1}{c}{\textsc{leaks}} & \multicolumn{1}{c}{\checkOK} \\
			\midrule
			{Andless} & 0.5 &  {\J{WakeLock}} & 13.1 & 1 & 1 & 1 & 1 \\
			{Apollo} & 24.5 & {\J{Camera}} & 176.4 & 2 & 2 & 4 & 2 \\
			{CheckCheck} & 4.5 &  {\J{WakeLock}} & 35.4 & 3 & 3 & 6 & 4 \\
			{Impeller} & 4.6 & {\J{WiFiLock}} & 25.9 & 4 & 3 & 1 & 1 \\
			{Jane} & 42.7 & {\J{LocationListener}}  & 280.5 & 2 & 2 & 0 & 0 \\
			{MiguMusic} & 13.0 & {\J{MediaPlayer}}  & 121.6 & 8 & 8 & 12 & 5 \\
			{PicsArt} & 17.6 &  {\J{WakeLock}} & 92.7 & 2 & 2 & 2 & 2 \\
			{QRScan} & 24.5 &  {\J{Camera}} & 144.1 & 4 & 4 & 6 & 3 \\
			{Runnerup} & 37.2 &  {\J{LocationListener}} & 232.9 & 13 & 11 & 8 & 5 \\
			{Shopsavvy} & 5.3 & {\J{LocationListener}}  & 41.3 & 8 & 7 & 5 & 3 \\
			{SimSimi} & 3.9 &  {\J{WakeLock}} & 35.6 & 7 & 6 & 11 & 4 \\
			{SuperTorch} & 3.8 & {\J{Camera}} & 24.5 & 3 & 3 & 1 & 1 \\
			{TigerMap} & 3.7 & {\J{LocationListener}}  & 36.7 & 2 & 2 & 6 & 2 \\
			{Utorch} & 1.0 & {\J{Camera}} & 20.5 & 2 & 2 & 1 & 1 \\
			{Vplayer} & 6.6 & {\J{MediaPlayer}}  & 49.8 & 7 & 5 & 7 & 5 \\
			{WeatherPro} & 5.0 & {\J{LocationListener}}  & 55.8 & 8 & 7 & 12 & 5 \\
			{Yelp} & 14.3 & {\J{LocationListener}}  & 156.4 & 2 & 2 & 2 & 2 \\
			\midrule
			\textsc{average} & 12.5 & & 90.8 \\
			\textsc{total} & 212.7 & & 1543.2 & 78 & 70 &	85 &	46 \\
			\bottomrule 
		\end{tabular}
    \end{adjustwidth}
    \caption{Comparison of \toolname's and Relda2's leak detection capabilities.
			For every \textsc{app} used in the comparison, the table
			reports its size \textsc{kloc} in thousands of lines of code,
			and the resource analyzed for leaks.
			Then, it reports \toolname's running \textsc{time} in seconds,
			the number of \textsc{leaks} \toolname detected, and how many
			of these we definitely confirmed as true positives by manual analysis (\checkOK).
			These results are compared to the number of \textsc{leaks} detected by Relda2,
			and how many of these were true positives (\checkOK) according to the experiments reported in \cite{Relda2TSE16}.
			The two bottom rows report the \textsc{average} (mean) and
			\textsc{total} for all apps/resources.
		}
		\label{tab:relda2}
	\end{table*}
	
	\begin{table*}[!tb]
		\centering
		\def\extraspace{\\[3pt]}
		\small
      \begin{adjustwidth}{-18mm}{-5mm}
		\begin{tabular}{lrl rrrrr rrr}
			\toprule
			& & & \multicolumn{5}{c}{\toolname} & \multicolumn{3}{c}{RelFix} \\
			\cmidrule(lr){4-8} \cmidrule(lr){9-11}
			\multicolumn{1}{c}{\textsc{app}} & \multicolumn{1}{c}{\textsc{kloc}} & \multicolumn{1}{c}{\textsc{resource}} & \multicolumn{1}{c}{\textsc{time} (s)} &
			\multicolumn{1}{c}{\textsc{leaks}} & \multicolumn{1}{c}{\checkOK} &
			\multicolumn{1}{c}{\textsc{fixed}} & \multicolumn{1}{c}{\textsc{invalid}} %
			& \multicolumn{1}{c}{\textsc{leaks}} & \multicolumn{1}{c}{\checkOK} &
			\multicolumn{1}{c}{\textsc{fixed}} \\
			\midrule
			\multirow{2}{*}{BlueChat} & \multirow{2}{*}{13.1} & {\J{MediaPlayer}} & 93.9 & 3 & 2 & 3 & 0 & 1 & 0 & 1  \\
			&  & \J{WakeLock} & 111.7 & 2 & 2 & 2 & 0 & 2 & 1 & 2 \extraspace
			\multirow{2}{*}{FooCam} & \multirow{2}{*}{14.7} & \J{Camera} & 152.7 & 1 & 1 & 1 & 0 & 3 & 1 & 3 \\
			&  & {\J{MediaPlayer}} & 124.5 & 0 & -- & 0 & 0  & 1 & 1 &  1 \extraspace
			GetBackGPS & 21.4 & {\J{LocationListener}} & 153.4 & 2 & 2 & 2 & 1 & 3 & 3 & 3 \extraspace
			SuperTorch & 3.8 & {\J{Camera}} & 24.5 & 3 & 2 & 3 & 1 & 1 & 1 & 1 \\
			\midrule
			\textsc{average} & 50.2 & & 110.1 \\
			\textsc{total} & 200.6 & & 660.7 & 11 & 9 & 11 & 2 & 11 & 7 & 11 \\
			\bottomrule
		\end{tabular}
    \end{adjustwidth}
    \caption{Comparison of \toolname's and RelFix's leak repair capabilities (on non-aliasing resources).
			For every \textsc{app} used in the comparison, the table
			reports its size \textsc{kloc} in thousands of lines of code,
			and the resources whose leaks are repaired.
			Then, it reports \toolname's running \textsc{time} in seconds,
			the number of leaks \textsc{detected} and \textsc{fixed} by \toolname, how many
			of these leaks we could conclusively classify as real leaks (\checkOK, true positives),
			and the number of fixes that \toolname classified as \textsc{invalid} (that is, they
			failed validation).
			These results are compared to the number of leaks \textsc{detected} and \textsc{fixed} by Relfix,
			and how many of these are considered real leaks (\checkOK)
			according to the experiments reported in \cite{Relfix}.
			The two bottom rows report the \textsc{average} (mean) and
			\textsc{total} for all apps/resources.
		}
		\label{tab:relfix}
	\end{table*}

	\subsubsection{RQ4: Comparison with Other Tools}
	\label{sec:answer-RQ4}
	\label{sec:comparison}
	
	\fakepar{Comparison with Relda2.}
	\autoref{tab:relda2} compares the leak detection capabilities of
	\toolname and Relda2 on the 17 apps used in the latter's experiments~\cite{Relda2TSE16}---%
	which only target non-aliasing resources.
	Relda2 reports more leaks than \toolname, but \toolname's precision (90\% $= 70/78$)
	is much higher than Relda2's (54\% $= 46/81$), and hence \toolname reports
	several more \emph{true} leaks (70 vs.\ 46).
	The difference between the two tools varies considerably with the app.
	On 4 apps (Andless, PicsArt, Vplayer, and Yelp)
	both tools detect the same number of leaks with the same (high) precision;
	even though we cannot verify this conjecture,
	it is quite possible that exactly the same leaks are detected by both tools
	in these cases.
	On 2 apps (Impeller and CheckCheck),
	neither tool is strictly better:
	when \toolname outperforms Relda2 in number of confirmed detected leaks
	(app Impeller) it also achieves a lower precision;
	when \toolname outperforms Relda2 in precision (app CheckCheck) it
	also finds one less correct leak.
	On the remaining 11 apps (Apollo, Jane, MiguMusic, QRScab, Runnerup, Shopsavvy, SimSimi, SuperTorch, TigerMap, Utorch, and WeatherPro)
	\toolname is at least as good as Relda2
	in both number of confirmed detected leaks and precision,
	and strictly better in detected leaks, precision, or both.
	We cannot directly compare \toolname's and Relda2's running times, since we
	could not run them on the same hardware, but we notice that the running times
	reported in \cite{Relda2TSE16} are in the ballpark of \toolname's.
	In all, \toolname's leak detection capabilities often outperforms Relda2's
	on the non-aliasing resources that we currently focus on.
	
	\fakepar{Comparison with RelFix.}
	\autoref{tab:relfix} compares the fixing capabilities of
	\toolname and RelFix (which uses Relda2 for leak detection)
	on the 4 non-aliasing resources used in the latter's experiments~\cite{Relfix}.
	Here too \toolname generally appears more effective than RelFix:
	conservatively assuming that all fixes reported by \cite{Relfix}
	are genuine and ``safe'',\footnote{RelFix's paper~\cite{Relfix} does not explicitly discuss possible fix validation errors.
	}
	\toolname has higher precision (82\%$= 9/11$ vs. 64\%$= 7/11$) and
	fixes at least as many confirmed leaks (even after discarding those
	that fail validation).
	
	\toolname generates small patches, each consisting of just 11 bytecode instructions
	on average, which corresponds to an average 0.017\% increase in size of a patched app.
	This approach leads to much smaller patches than RelFix's, whose
	patches also include instrumentation to detect the leaks on which
	the fixes depend to function correctly.
	As a result, the average increase in size introduced by a RelFix patch is 0.3\%
	in terms of bytecode instructions~\cite{Relfix},
	which is one order of magnitude larger than \toolname's.
	
	As explained in \autoref{subjects-comparison}, all results in this section
	are subject to the limitation that a direct comparison with Relda2 and RelFix
	was not possible.
	Despite this limitation, the comparison collected enough evidence to indicate
	that \toolname's analysis is often more thorough
	and more precise than the other tools'.

	\begin{result}
		On non-aliasing resources, \toolname is usually \\
		more effective and precise than 
		other techniques for the automated detection and repair of
		Android resource leaks.
	\end{result}
	
	\begin{table}
		\centering
		\def\extraspace{\\[1pt]}
		\normalsize
		\begin{tabular}{r rrr rr}
			\toprule
			& \multicolumn{3}{c}{\leaks}  & \multicolumn{2}{c}{Relda2/RelFix} \\
			\cmidrule(rl){2-4} \cmidrule(rl){5-6}
			$D$ & \multicolumn{1}{c}{\textsc{time} (s)}  & \multicolumn{1}{c}{\textsc{fixed}} & \multicolumn{1}{c}{\textsc{missed}} 
			& \multicolumn{1}{c}{\textsc{fixed}} & \multicolumn{1}{c}{\textsc{missed}} \\
			\midrule
			1 & 116.7 & 40 & 10 & 75  & 14 \\
			2 & 184.9 & 50 & 0  & 84  & 5  \\
			3 & 320.2 & 50 & 0  & 89  & 0  \\
			4 & 501.3 & 50 & 0  & 89  & 0  \\
			5 & 907.8 & 50 & 0  & 89  & 0  \\
			6 & 1894.3 & 50 & 0 & 89  & 0  \\
			\bottomrule 
		\end{tabular}
		\caption{Results of running \toolname with different unrolling depths
			on the \leaks benchmark and on the apps used in the comparison with Relda2 and RelFix.
			For each unrolling depth $D$,
			the table reports the \textsc{average} (mean, per app-resource)
			running \textsc{time} of
			\toolname on all leaks of non-aliasing resources;
			and
			the total number of \textsc{fixed leaks} and 
			\textsc{missed leaks} (not detected, and hence not fixed).
			The row with $D=3$ corresponds to the \leaks data in \autoref{tab:results},
			and the Relda2/RelFix data in Tables~\ref{tab:relda2}--\ref{tab:relfix}.
		}
		
		\label{tab:results-unrolling}
	\end{table}
		
	\subsubsection{RQ5: Unrolling}
	\label{sec:answer-RQ5}
	
	In the experiments reported so far, \toolname ran with the unrolling depth parameter $D = 3$, which is the default.
	\autoref{tab:results-unrolling} summarizes the key data about experiments on the same apps
	but using different values of $D$.
	These results indicate that a value of $D \geq 3$ is required for soundness:
	\toolname running with $D=1$ missed 24~leaks (10 in \leaks, 
	and 14 in the apps used by Relda2/RelFix);
	with $D=2$ it missed 5 leaks (all in the apps used by Relda2/RelFix).
	All missed leaks only occur with resources that are acquired multiple times---that is they
	affect \emph{reentrant} resources: \J{WakeLock} and \J{WifiLock}
	in apps CallMeter, CSipSimple, IRCCloud (from \leaks),
	CheckCheck, Impeller, PicsArt, SimSimi (from the comparison with Relda2),
	and BlueChat (from the comparison with RelFix).
	
	Is $D=3$ also sufficient for soundness?
	While we cannot formally prove it,
	the experiments suggest this is the case: increasing $D$ to larger values does not find new
	leaks but only increases the running time.
	Unsurprisingly,
	the running time grows conspicuously with the value of $D$,
	since a larger unrolling depth determines a combinatorial increase in the number of possible
	paths.
	Thus, for the analyzed apps, 
	the default $D=3$ is the empirically optimal value:
	it achieves soundness without unnecessarily increasing the running time.
	It is possible this result does not generalize to apps with more complex reentrant resource
	management; however, \toolname always offers the possibility of increasing $D$
	until its analysis is sufficiently exhaustive.

	\begin{result}
		\toolname's analysis is sound provided it unrolls callback loops a
		sufficient number of times (thrice in the experiments).
	\end{result}
	
	\subsection{Aliasing Resources}
	\label{sec:aliasing}
	
	We repeatedly remarked that \toolname's current implementation
	is effective only on non-aliasing resources;
	it remains applicable to aliasing resources, but it is bound to generate a large number of false positives.
	In order to get a clearer picture of \toolname's limitations in the presence of aliasing,
	this section reports some additional experiments on aliasing resources.
	It remains that our contributions focus on non-aliasing resources; an adequate support of aliasing
	belongs to future work.
	
	\begin{table}[!hbt]
		\centering
		\setlength{\tabcolsep}{4pt}
		\begin{tabular}{llrrrr}
			\toprule
			\multicolumn{1}{c}{\textsc{tool}} & \multicolumn{1}{c}{\textsc{subjects}} & \multicolumn{1}{c}{\textsc{resources}} & \multicolumn{1}{c}{\textsc{leaks}} & \multicolumn{1}{c}{\textsc{\checkOK}}  & \multicolumn{1}{c}{\textsc{precision}} \\
			\midrule
			Relda2 & \cite{Relfix} & 4 & 108 & 24 & 22\% \\
			\cmidrule(lr){1-1}
			\toolname & \leaks & 6 & 273 & 86 & 32\% \\
			\bottomrule
		\end{tabular}
		\caption{Detection of leaks of \emph{aliasing} resources by Relda2 (according to the experiments reported in \cite{Relfix}) and \toolname (on the programs in \leaks). The table reports the number of aliasing \textsc{resources} involved in leaks, the reported \textsc{leaks}, how many of them are confirmed true leaks \checkOK, and the corresponding precision $\checkOK/\textsc{leaks}$.}
		\label{tab:aliasing}
	\end{table}
	
	\autoref{tab:aliasing} shows the behavior of \toolname
	on leaks of the 6 aliasing resources in \leaks apps.
	\toolname reported a total of 273 leaks, but we could only confirm about one third of them as
	real leaks (true positives). This is a lower bound on \toolname's precision on aliasing resources,
	as it's possible that more reported leaks are real but we could not conclusively confirm them
	as such because they affect obfuscated apps. Nevertheless, it's clear the precision is much lower
	than on non-aliasing resources---as we expected.
	
	For comparison,
	\autoref{tab:aliasing} also reports the performance of Relda2 according
	to the experiments reported in \cite{Relfix} on 
	4 aliasing resources (\autoref{sec:comparison} discusses the data from the same source but involving non-aliasing
	resources).
	Relda2 reported a total of 108 leaks, but only 22\% of them are real leaks according to~\cite{Relfix}.
	Even though the experiments with \toolname and with Relda2 are not directly
	comparable, it's clear both tools achieve a low precision on aliasing resources---arguably low
	to the point that practical usefulness is severely reduced.
	
	\subsection{Threats to Validity}
	\label{sec:threats}
	
	The main threats to the validity of our empirical evaluation come from the fact that we analyzed Android apps
	in \emph{bytecode} format; furthermore, some of these apps' bytecode was only available in \emph{obfuscated} form.
	In these cases, we were not able to inspect in detail how the fixes modified the original programs; 
	we could not always match with absolute certainty the leaks and fixes listed in \leaks with the fixes produced by \toolname;
	and we could not run systematic testing of the automatically fixed apps.
	This threat was significantly mitigated by other sources of evidence that \toolname indeed produces fixes that are correct
	and do not alter program behavior except for removing the source of leaks:
	first, the manual inspections we could carry out on the apps that are not obfuscated confirmed in all cases our expectations;
	second, \toolname's analysis is generally \emph{sound}, and hence it should detect all leaks
	(modulo bugs in our implementation of \toolname);
	third, running the fixed apps for significant periods of time did not show any apparent change in their behavior.
	
	As remarked in \autoref{subjects-comparison},
	we could only perform an indirect comparison with
	Relda2/RelFix (the only other fully automated approach
	for the repair of Android resource leaks that is currently available)
	since neither the tools nor details of their experiments
	other than those summarized in their publications~\cite{Relda2TSE16,Relfix}
	are available.
	To mitigate the ensuing threats, we analyzed app versions that were available
	around the time when the Relda2/RelFix experiments were conducted,
	and we excluded from the comparison with \toolname
	measures that require experimental repetition (such as running time).
	While it is still possible that measures such as precision were
	assessed differently than how we did, these should be minor differences that do not invalidate
	the high-level results of the comparison.
	
	Our evaluation did not assess the acceptability of fixes from a programmer's perspective.
	Since \toolname works on bytecode, its fixes may not be easily accessible by developers familiar only with the source code.
	Nonetheless, fixes produced by \toolname are succinct and correct by construction, which is usually conducive to readability and acceptability.
	As future work, one could implement \toolname's approach at the level of source code,
	so as to provide immediate feedback to programmers as they develop an Android app.
	\toolname in its current form could instead be easily integrated in an automated checking system for Android apps---for example,
	within app stores.
	
	We didn't formally prove the \emph{soundness} or \emph{precision}
	of \toolname's analysis,
	nor that our implementation is free from bugs.
	Since \toolname is implemented on top of \textsc{AndroGuard}~\cite{AndroGuard} and \textsc{Apktool}~\cite{Apktool}
	(\autoref{sec:implementation}),
	any bugs or limitations of these tools may affect \toolname's analysis.
	In particular, \textsc{AndroGuard} cannot currently analyze native
	code\footnote{\url{https://github.com/androguard/androguard/issues/566\#issuecomment-431090708}}---a
	common limitation of static analysis.
	By and large, however, these tools' support of Android is quite extensive,
	and hence any current limitations are unlikely to significantly impact the soundness
	of the results obtained in \toolname's experimental evaluation.
	
	Nonetheless, we analyzed in detail the features of \toolname's analysis
	both theoretically (\autoref{sec:limitations}) and empirically (\autoref{sec:experiments}).
	The empirical evaluation corroborates the evidence that \toolname
	is indeed sound (for sufficiently large unrolling depth $D$),
	and that \emph{aliasing} is the primary source of imprecision.
	In future work, we plan to equip \toolname with alias analysis~\cite{KelloggSSE2021},
	in order to boost its precision on the aliasing resources that currently lead to many false positives.
	
	\leaks offers a diverse collection of widely-used apps and leaked resources,
	which we further extended with apps used in Relda2/RelFix's evaluations.
	In our experiments, we used all apps and resources in \leaks and in Relda2/RelFix's evaluations
	that \toolname can analyze with precision.
	This helps to generalize the results of our evaluation, and
	led to finding numerous leaks not included in \leaks nor found by Relda2/RelFix.
	Further experiments in this area would greatly benefit from
	extending curated collections of leaks and repairs like \leaks.
	Our replication package is a contribution in this direction.
	
	\section{Related Work}
	\label{sec:relatedwork}
	
	\begin{table*}[!htb]
		\centering
		\def\extraspace{\\[3pt]}
		\small
      \begin{adjustwidth}{-20mm}{-5mm}
		\begin{tabular}{lllcccc}
			\toprule
			\multicolumn{1}{c}{\textsc{tool}} & \multicolumn{1}{c}{\textsc{analysis}} &
			\multicolumn{1}{c}{\textsc{leaks}} & \multicolumn{1}{c}{\textsc{automated}} &
			\multicolumn{1}{c}{\textsc{reentrant}} & 
			\multicolumn{1}{c}{\textsc{soundness}} & 
			\multicolumn{1}{c}{\textsc{repair}} \\
			\midrule
			FindBugs \cite{FindBugs}& static & Java resources & full & No & No & 
			No \\
			EnergyTest \cite{FSE14}& dynamic & energy & partial & No & No & 
			No \\
			LeakCanary \cite{LeakCanary}& dynamic & memory & full & No & No &
			No \\
			Android Studio \cite{AndroidStudioMonitor}& dynamic & memory & full & No & No & 
			No \\                                                            
			FunesDroid \cite{AmalfitanoRTF20}& dynamic & memory & partial & No & No & 
			No \\                                                                             
			Sentinel \cite{Sentinel,WuWR18}& static$+$dynamic & sensor & partial & Yes & No & 
			No \\
			Relda2/RelFix \cite{Relda2TSE16,Relfix}& static & Android resources & full & No & No & 
			(Yes) \\
			EnergyPatch \cite{EnergyPatchTSE18,MobiSoft16}& static$+$dynamic & energy & full & No & No & 
			(Yes) \\
			\cmidrule(lr){1-1}
			\toolname & static  & Android resources & full  & Yes & Yes & 
			Yes \\
			\bottomrule
		\end{tabular}
    \end{adjustwidth}
    \caption{Comparison of tools that detect and repair leaks in Android apps. For each tool, the table report the kinds of code \textsc{analysis} it performs (static or dynamic), the kinds of \textsc{leaks} it primarily targets, whether it is fully or partially \textsc{automated}, whether it models \textsc{reentrant} behavior, whether its detection is \textsc{sound}, and whether it can also generate \textsc{repair}s of the detected leaks. Entries (Yes) in column \textsc{repair} denote fixes that release any leaking resource at the very end of an app's lifecycle, without following the resource-specific Android guidelines. }
		\label{tab:related-work}
	\end{table*}
	
	\subsection{Automated Program Repair}
	\label{sec:related-APR}
	\toolname is a form of automated program repair (APR) targeting a specific kind of bugs (resource leaks) and programs (Android apps).
	The bulk of ``classic'' APR research~\cite{GazzolaAPRSurvey,monperrus-survey,genprog,defects4j-repair} usually targets general-purpose techniques,
	which are applicable in principle to any kinds of program and behavioral bugs.
	The majority of these techniques are based on dynamic analysis---that is, they rely on \emph{tests} to define expected behavior, to detect and localize errors~\cite{SketchFix,ChenPeiFuria},
	and to validate the generated fixes~\cite{LongR15,SimFix,ELIXIR}.
	General-purpose APR completely based on static analysis is less common~\cite{LogozzoB12StaticAPR,LogozzoM13StaticAPR,GaoXMZYZXM15StaticAPR},
	primarily because tests are more widely available in general-purpose applications,
	whereas achieving a high precision with static analysis is challenging for the same kind of applications and properties.
	
	\subsection{Leak Analysis}
	Whereas \toolname is one of only two fully automated approaches
	for \emph{fixing Android resource} leaks (the other is RelFix, discussed below and in \autoref{sec:answer-RQ4}),
	\emph{detection} of leaks and other defects is more widely studied and
	has used a broad range of techniques---from static analysis to testing.
	\autoref{tab:related-work} outlines the features of the main related approaches, which we discuss in the rest of this section.

	\subsubsection{Static Analysis}
	\label{sec:related-static}
	Approaches based on static analysis build an \emph{abstraction} of a program's behavior,
	which can be searched exhaustively for leaks.
	Since Android apps run on mobile devices, they
	are prone to defects such as privacy leaks~\cite{GiblerCEC12}, permission misuses~\cite{AnadroidMalware}, and other security vulnerabilities~\cite{secVul_ndss/ZhouJ13, SA_survey2020}
	that are less prominent (or have less impact) in traditional ``desktop'' applications.
	In such specialized domains, where \emph{soundness} of analysis is paramount,
	static analysis is widely applied---for example to perform taint analysis~\cite{taintanalysisLuoBS19}
	and other kinds of control-flow based analyses~\cite{StaticAnalysisSurveyLiBPRBOKT17,FlowDroid_PLDI14}.
	Indeed, there has been plenty of work that applied static analysis to analyze resource management in Java~\cite{WeimerN04,DilligDYC08},
	including detecting resource leaks~\cite{TorlakC10, KelloggSSE2021}.
	
	Whereas some of these contributions could be useful also to analyze mobile apps written in Java,
	in order to do so the techniques should first be extended to support the peculiarities of the Android programming model---in particular, its event-driven control flow.
	
	General-purposes static analyzers for Java like FindBugs~\cite{FindBugs},
	are also capable of detecting a variety of 
	issues in common Java resources (e.g., files);
	however, FindBugs is neither sound nor very precise~\cite{DBLP:conf/msr/VetroTM10},
	as it is based on heuristics and pattern-matching that
	are primarily geared towards detecting stylistic issues and
	code smells.
	
	\subsubsection{Dynamic Analysis}
	Many approaches use dynamic analysis (testing), and hence are not sound
	(they may miss leaks).
	
	One example 
	is a test-gen\-er\-a\-tion framework 
	capable of building inputs exposing resource leaks that lead to energy inefficiencies~\cite{FSE14}.
	Since it targets energy efficiency, the framework consists of a hybrid setup that includes hardware to physically measure energy consumption, which makes it less practical to deploy (hence, \autoref{tab:related-work} classifies it as ``partial'' automation).
	Its measurements are then combined with more traditional software metrics to generate testing oracles for energy leak detection.
	The framework's generated tests are sequences of UI events
	that trigger energy leaks or other inefficiencies exposed by the oracles.
	As it is usual for test-case generation,
	\cite{FSE14}'s framework is based on heuristics and statistical assumptions about energy consumption patterns,
	and hence its detection capabilities are not exhaustive (i.e., not sound).
	
	Other tools~\cite{Sentinel,WuWR18,LeakCanary,AndroidStudioMonitor,AmalfitanoRTF20,YanYR13} exist that
	use tests to detect resource leaks---such as sensor leaks~\cite{Sentinel,WuWR18,YanYR13} and memory leaks~\cite{LeakCanary,AmalfitanoRTF20}.
	A key idea underlying these approaches is to combine runtime resource profiling~\cite{AndroidStudioMonitor} and search-based test-case generation
	looking for inputs that expose leaks.
	
	LeakCanary~\cite{LeakCanary} and the Android Studio Monitor~\cite{AndroidStudioMonitor} are two of the most popular tools used by developer to detect memory leaks as they have high \emph{precision} and are fully automated.
	Like all approaches based on testing, they are unsound in general, and their
	capabilities strongly depend on the quality of the tests that are provided.
	
	FunesDroid \cite{AmalfitanoRTF20} generates inputs that correspond to pre-defined user interactions---like rotating the screen---which can trigger memory leaks when executed in certain activities. Besides being unsound like all testing techniques, FunesDroid's testing capabilities are also limited by the kinds of interactions that it supports. Furthermore,
	FunesDroid's test generation capabilities are not fully automated for leak detection: while the tool provides
	user interactions as inputs, one still needs to provide tests that run the app where leaks are being detected.
	
	Sentinel~\cite{Sentinel,WuWR18}, an approach based on generating GUI events for testing, models resource usage as context-free languages, and hence it is one of the few leak analysis techniques that is capable of fully modeling reentrant resources. The tool first performs a static analysis of app code to build a model
	that maps GUI events to callback methods that affect sensor behavior. Then, it traverses the model to enumerate paths, which in turn are used to generate test cases. The last step (from paths to actual test inputs)
	requires users to manually come up with suitable inputs that match the abstract paths;
	therefore, the Sentinel approach is not fully automated.
	
	\subsection{Leak Repair}
	\label{sec:related-leaks}
	The amount of work on \emph{detecting} various kinds of  leaks~\cite{EnergyPatchTSE18,FSE14,Relda2TSE16,DynamicRLA_LiuXCL14, LeakEmpericalStudy}
	and the recent publication of the \leaks curated collection of leaks~\cite{droidleaks}
	indicate that leak detection is considered a practically important problem in Android programming.
	In this section, we discuss in greater detail
	two approaches that are also capable, like \toolname, of \emph{fixing}
	the Android leaks they detect.
	
	\subsubsection{Relda2 and RelFix}
	Relda2~\cite{Relda2TSE16} combines flow-insensitive and flow-sensitive static analyses
	to compute resource \emph{summaries}:
	abstract representations of each method's resource usage,
	which can be combined to perform leak detection across different procedures and callbacks.
	Since Relda2 approximates loops in an activity's lifecycle
	by abstracting away some of the flow information (even in its ``flow-sensitive'' analysis),
	and does not accurately track nested resource acquisitions,
	its analysis is generally unsound (leaks may go undetected) and imprecise (spurious errors may be reported). In contrast, \toolname performs a more thorough and precise inter-procedural analysis by considering all possible callback sequences. \toolname even allows users to set the \emph{unrolling depth} $D$ (affecting the maximum length of analyzed callback sequences),
	which is a means of trading off soundness (i.e., how thorough the analysis is)
	with running time (i.e., how long/how many computational resources the analysis takes).
	\toolname's modeling of resources is more detailed than Relda2's also because it supports
	a validation step (\autoref{sec:validation}),
	which can detect inconsistencies between a resource's recommended usage guidelines
	and how it is actually used by an app.
	After building a control-flow model of resource usages,
	Relda2 uses an off-the-shelf model checker to analyze it.
	In contrast, \toolname uses a custom automata-theoretic analysis algorithm
	to search for leaks on resource-flow graphs, which may contribute to more precise
	and scalable results.
	
	RelFix~\cite{Relfix} can patch resource leaks in Android apps that
	have been detected by Relda2. It applies patches at the level of Dalvik bytecode, whereas \toolname does so at the level Smali (a human-readable format of Dalvik),
	which makes \toolname's output more accessible to human programmers.
	\autoref{sec:comparison} discussed a detailed comparison between \toolname and Relda2/RelFix in terms of precision and patch sizes. Another significant difference is how
	they build fixes: RelFix follows the simple approach of releasing resources in the very last callback of an activity's lifecycle,
	whereas \toolname builds fixes that adhere to Android's recommended guidelines.
	
	\subsubsection{EnergyPatch}
	EnergyPatch~\cite{EnergyPatchTSE18,MobiSoft16} is another approach for leak detection based on static techniques where resource usage are modeled as regular expressions.
	EnergyPatch uses abstract interpretation to compute an over-approximation of an app's energy-relevant behavior;
	then, it performs symbolic execution to detect which abstract leaking behaviors are false positives and which are executable
	(i.e., correspond to a real resource leak);
	therefore, its analysis is hybrid as it combines static and dynamic techniques.
	For each executable leaking behavior, symbolic execution can also generate a concrete program input that triggers
	the energy-leak bug. 
	EnergyPatch targets a different kind of resource leaks (energy-consumption related, which lead to a wasteful usage of a mobile device's energy resources) than \toolname.
	Since EnergyPatch only analyzes simple paths (i.e., without loops) in the callback graph,
	its analysis may be unsound (especially for reentrant resources).

	While EnergyPatch focuses on leak detection,
	it also offers a simple technique for generating fixes,
	which simply releases all resources in the very last callback of an activity's lifecycle.
	As we discussed in \autoref{sec:fixgen},
	this approach is sometimes impractical
	because it may conflict with some of Android programming's best practices. In contrast, \toolname releases the resource aggressively, in the earliest callback, since our validation step later can filter out patches that result in \emph{use-after-release} issues. 
	
	\section{Conclusions and Future Work}
	\label{sec:conclusions}
	This paper presented \toolname: a technique and tool to detect and automatically fix resource leaks in Android apps.
	\toolname is based on succinct static abstractions of an app's control-flow;
	therefore, its analysis is sound and its fixes are correct by construction.
	Its main limitation is that its analysis tends to generate false positives on resources that are frequently \emph{aliased} within the same app.
	In practice, this means that \toolname's is currently primarily designed for the numerous Android resources that are not subject to aliasing.
	On these resources, we demonstrated \toolname's effectiveness and scalability.
	Extending \toolname's approach with aliasing information is an
	interesting and natural direction for future work.

\bibliographystyle{plain}
\bibliography{plumbdroid}

\end{document}